\documentclass[10pt,journal,compsoc]{IEEEtran}
\ifCLASSOPTIONcompsoc
\usepackage[nocompress]{cite}
\else
\usepackage{cite}
\fi

\usepackage[]{hyperref}

\usepackage{enumitem}
\usepackage{tightenum}
\usepackage{listings}
\usepackage{tabularx}
\usepackage{diagbox}
\usepackage{setspace}
\usepackage{epsfig}

\usepackage{url}

\usepackage{subcaption}
\usepackage[ruled]{algorithm2e}

\usepackage{booktabs}
\usepackage{multirow}
\usepackage{makecell}
\usepackage{tikz}
\usepackage{wrapfig}

\newcommand{\songlh}[1]{{\color{red} \sf (LS: #1)}}
\newcommand{\yiying}[1]{{\color{blue} \sf (YZ: #1)}}
\newcommand{\mengting}[1] {{\color{brown}(MT: #1)}}

\newcommand{\x}{$\times$}

\newcommand{\ignore}[1]{}

\newcommand{\beforecaption}{\vspace{-.15cm}\begin{spacing}{0.85}}
\newcommand{\aftercaption}{\vspace{-.15cm}\end{spacing}}

\newcommand{\mycaption}[3]{\beforecaption\caption{\label{#1}{#2} #3}\aftercaption}

\newcommand{\eg}{\textit{e.g.}}
\newcommand{\ie}{\textit{i.e.}}
\newcommand{\etal}{\textit{et al.}}

\newcommand{\bolditalicparagraph}[1]{\noindent{\textit{\textbf{#1}}}}

\newcommand{\italicparagraph}[1]{\noindent\underline{\textit{#1}}}

\newcommand{\boldunderparagraph}[1]{\noindent\underline{\textbf{#1}}}

\newcounter{insight}
\newcounter{suggestion}
\usepackage{listings, xcolor}

\definecolor{verylightgray}{rgb}{.97,.97,.97}

\lstdefinelanguage{Solidity}{
	keywords=[1]{anonymous, assembly, assert, balance, break, call, callcode, case, catch, class, constant, continue, constructor, contract, debugger, default, delegatecall, delete, do, else, emit, event, experimental, export, external, false, finally, for, function, gas, if, implements, import, in, indexed, instanceof, interface, internal, is, length, library, log0, log1, log2, log3, log4, memory, calldata, modifier, new, payable, pragma, private, protected, public, pure, push, require, return, returns, revert, selfdestruct, send, solidity, storage, struct, suicide, super, switch, then, this, throw, transfer, true, try, typeof, using, value, view, while, with, addmod, ecrecover, keccak256, mulmod, ripemd160, sha256, sha3}, 
	keywordstyle=[1]\color{blue}\bfseries,
	keywords=[2]{address, bool, byte, bytes, bytes1, bytes2, bytes3, bytes4, bytes5, bytes6, bytes7, bytes8, bytes9, bytes10, bytes11, bytes12, bytes13, bytes14, bytes15, bytes16, bytes17, bytes18, bytes19, bytes20, bytes21, bytes22, bytes23, bytes24, bytes25, bytes26, bytes27, bytes28, bytes29, bytes30, bytes31, bytes32, enum, int, int8, int16, int24, int32, int40, int48, int56, int64, int72, int80, int88, int96, int104, int112, int120, int128, int136, int144, int152, int160, int168, int176, int184, int192, int200, int208, int216, int224, int232, int240, int248, int256, mapping, string, uint, uint8, uint16, uint24, uint32, uint40, uint48, uint56, uint64, uint72, uint80, uint88, uint96, uint104, uint112, uint120, uint128, uint136, uint144, uint152, uint160, uint168, uint176, uint184, uint192, uint200, uint208, uint216, uint224, uint232, uint240, uint248, uint256, var, void, ether, finney, szabo, wei, days, hours, minutes, seconds, weeks, years},	
	keywordstyle=[2]\color{teal}\bfseries,
	keywords=[3]{block, blockhash, coinbase, difficulty, gaslimit, number, timestamp, msg,  gas, sender, sig, value, now, tx, gasprice, origin},	
	keywordstyle=[3]\color{violet}\bfseries,
	identifierstyle=\color{black},
	sensitive=true,
	comment=[l]{//},
	morecomment=[s]{/*}{*/},
	commentstyle=\color{gray}\ttfamily,
	stringstyle=\color{red}\ttfamily,
	morestring=[b]',
	morestring=[b]"
}

\usepackage{amsmath,amsfonts}
\usepackage{algorithmic}
\usepackage{stfloats}
\usepackage{graphicx}
\usepackage{textcomp}
\usepackage{xcolor}

\usepackage[most]{tcolorbox}

\def\BibTeX{{\rm B\kern-.05em{\sc i\kern-.025em b}\kern-.08em
    T\kern-.1667em\lower.7ex\hbox{E}\kern-.125emX}}

\begin{document}

\title{How to Save My Gas Fees: 
Understanding and Detecting Real-World Gas Issues in Solidity Programs}

\author{Mengting He, Shihao Xia, Boqin Qin, Nobuko Yoshida, Tingting Yu, Yiying Zhang, Linhai Song 
\thanks{Mengting He, Shihao Xia, and Linhai Song are with the Pennsylvania State University, PA, USA, 16802. E-mails: mvh6224@psu.edu, szx5097@psu.edu, and songlh@ist.psu.edu.}
\thanks{Boqin Qin is with China Telecom Cloud Technology Co., Ltd, Beijing, China, 100083. E-mail: bobbqqin@gmail.com.}
\thanks{Nobuko Yoshida is with the University of Oxford, Oxford, UK, OX1 3QG. E-mail: nobuko.yoshida@cs.ox.ac.uk.}
\thanks{Tingting Yu is with the University of Connecticut, CT, USA, 06269. E-mail: tingting.yu@uconn.edu.}
\thanks{Yiying Zhang is with the University of California, San Diego, CA, USA, 92093. E-mail: yiying@ucsd.edu.}
\thanks{Linhai Song is the corresponding author}
}


\IEEEtitleabstractindextext{
		The execution of smart contracts on Ethereum, a public blockchain system, incurs a fee called \textit{gas fee} for its computation and data storage.
When programmers develop smart contracts (\eg, in the Solidity programming language), they could unknowingly write code snippets that unnecessarily cause more gas fees. These issues, or what we call \textit{gas wastes}, can lead to significant monetary losses for users. 
This paper takes the initiative in helping Ethereum users reduce their gas fees in two key steps. First, we conduct an empirical study on gas wastes in open-source Solidity programs
and Ethereum transaction traces.
Second, to validate our study findings, 
we develop a static tool called \textit{PeCatch} to effectively detect gas wastes in Solidity programs, and 
manually examine the Solidity compiler’s code to 
pinpoint implementation errors causing gas wastes. 
Overall, we make 11 insights and four suggestions, which can foster future tool development and programmer awareness, and fixing our detected bugs can save \$0.76 million in gas fees daily.

\begin{IEEEkeywords}
Smart Contracts, Bug Study, Static Bug Detection, Gas Optimization
\end{IEEEkeywords}}


\maketitle

\IEEEdisplaynontitleabstractindextext
\IEEEpeerreviewmaketitle

\section{Introduction}
\label{sec:intro}

\IEEEPARstart{A}{s} a public blockchain system, Ethereum enables the deployment 
and execution of smart contracts~\cite{eth-1,eth-2}, 
allowing developers to create sophisticated decentralized 
applications (dapps)~\cite{smart-contract}. These smart contracts are essential 
to Ethereum's digital economy, facilitating over 1 million daily transactions 
with a total volume exceeding \$4 billion\cite{eth-daily}. 
Additionally, several projects on Ethereum have reached impressive market values of surpassing \$1 billion~\cite{uniswap,opensea}.

Solidity, the official programming language of Ethereum~\cite{solidity-1}, 
resembles ECMAScript 
and simplifies smart contract development by concealing the complexities of Ethereum Virtual Machine (EVM) 
and integrating security features. It has gained immense popularity among 
smart-contract programmers~\cite{solidity-popular-1,solidity-popular-2}, with almost all Ethereum contracts written in Solidity~\cite{solidity-top-1:url}, adding over 1 million new contracts 
quarterly.

To safeguard Ethereum's computational resources from DoS attacks and
tax Ethereum transactions,
a fee called \textit{gas} is charged for executing each smart contract~\cite{gas-1:url,gas-2:url}. 
Besides correctness, performance, and security, gas is another crucial metric
for smart contracts on Ethereum. 
Various EVM operations incur different gas costs, and some may 
have a complicated gas computation formula.  
While Solidity hides the complexities of EVM, 
it also obscures how gas fees are charged for each piece of Solidity code. Consequently, Solidity programmers struggle to optimize gas usage, often resulting in gas-inefficient code (\ie, code that can be optimized to use less gas). 
Unfortunately, the Solidity compiler lacks sufficient optimizations for gas efficiency, leading to significant gas wastage on Ethereum.

Given the significant daily gas fee expenditure by Solidity programs on Ethereum~\cite{gastracker}, optimizing code for gas efficiency is paramount. 
To achieve this over-arching goal, the foundational step is to comprehend how gas inefficiency occurs 
in real Solidity programs and how such inefficiency could be avoided. Doing so can guide the 
focus of tool development and language improvements, which in turn could help Ethereum users save 
substantial amounts of money. 
In this paper, we define one gas-inefficient code site 
in a Solidity program
as one \textit{gas waste}.

While defects affecting correctness and 
performance have been extensively 
studied in traditional 
programs~\cite{manna1969correctness,bug-change,ldoctor,CARAMEL,PerfBug}, 
the methodologies 
and tools used to detect these bugs cannot be directly applied to gas wastes 
in Solidity programs due to their unique characteristics: 
the amount of gas consumption by a code segment 
is \emph{not proportional to its performance or execution outcome}. 
Moreover, 
the EVM’s unique non-register-based instructions,
intricate data-store system, and complex gas cost model mean that gas wastes have \emph{root causes and code patterns not found in other programming languages}.

To understand gas wastes in Solidity,
we take an empirical approach. We first study existing, 
patched gas wastes in real Solidity programs. 
We then analyze on-chain traces of Ethereum transactions to inspect
wastes that may go unnoticed by programmers and thus escape into production runs. 
Finally, 
we develop gas-waste detection tools that automatically identify previously undiscovered gas wastes in Solidity programs and manually capture bugs in the Solidity compiler 
to demonstrate the value of the study results. 
Our empirical approach reveals important characteristics of 
gas wastes in real life --- 
those gas wastes programmers are likely to make and cost much money, thereby directing future efforts in more effective directions.

We study 100 gas wastes, including 54 specific to Solidity, 
from five popular Solidity applications. We categorize these wastes 
based on the data store areas where the gas-inefficient code executes: 
\textit{stack}, where instruction operands and results sit on; 
\textit{memory}, a volatile and byte-addressable area; 
\textit{storage}, a persistent key-value data store; 
and \textit{calldata}, byte-addressable arrays used for sending data across contracts. 
Each of these areas has different gas cost models (\eg, storage consumes 50\x\ 
more gas than stack and memory) and uses various types of instructions, leading to different root causes and buggy code patterns for gas wastes. After inspecting
wastes in different groups, 
we find that the misuse of stack and storage are the primary 
and secondary causes of gas wastes,
that the Solidity compiler may cause gas wastes
when trading gas usage for contract reliability, and that programmers often 
write gas-inefficient code that does not properly 
utilize Solidity's unique language features.

Our on-chain trace analysis examines ten million Ethereum transactions 
over ten days, revealing a total gas fee of \$160 million. 
We focus primarily on frequently executed opcode sequences 
and those without any side effects after execution, 
aiming to collect gas wastes with substantial fixing rewards. 
After analyzing the characteristics of the collected wastes, 
we observe that contract programmers may incorrectly balance gas usage when exercising transactions of a contract 
and gas usage when deploying the contract,
and that many wastes 
on the stack are due to compiler implementation issues. 
Addressing bugs in the compiler, particularly those affecting opcode generation 
for stack manipulation, is essential to reduce gas consumption.

In total, our analysis of real-world gas wastes and online traces yields 
11 insights and four suggestions. Most of these insights (except Insight 4) 
have not been reported in previous literature. While some of them 
(Insight 4 and Suggestions 1, 3, and 4) overlap with findings 
from existing papers, our focus is on understanding how gas is wasted in practice 
and we provide real data to demonstrate how frequently they impact the real world 
and their monetary consequences, making the reporting of these findings still 
valuable in practical terms.


To identify gas wastes in Solidity programs,
we develop \textit{PeCatch}, a suite of six static checkers. 
We implement these checkers with careful consideration of Solidity’s unique store mechanisms 
and language features (\eg, \texttt{unchecked}, \texttt{calldata}).
We evaluate PeCatch on the latest versions of the five studied Solidity applications and four additional open-source Solidity projects. 
PeCatch detects a total of 302 previously unknown gas wastes 
in the benchmark programs, 
significantly more than existing techniques~\cite{slither,brandstatter2020characterizing,gassaver,grech2018madmax, llvm-analysis,solidity-optimizations},
while reporting zero false positives. 
Additionally, we pinpoint 14 bugs causing gas wastes 
in the Solidity compiler. 
Fixing the detected wastes and compiler bugs could save \$0.76 million every day. 
The effectiveness and accuracy of PeCatch, along with the substantial monetary impact of identified issues, underscore the value of our empirical study.

In sum, we make the following contributions.

\begin{itemize} 

    \item We conduct an empirical study on real-world gas wastes, deriving 11 key insights and proposing four recommendations for future language designers and programmers.   
    
\item We develop PeCatch, a tool much more effective in gas-waste detection than state-of-the-art techniques.

\item We uncover numerous issues in Solidity programs and the Solidity compiler, with a large monetary impact.

\end{itemize}

We have released our study results, source code of our bug detector,
and detailed experimental results, all of which  
can be found at \href{https://github.com/PeCatch-Artifact/PeCatch-Artifact}{\url{https://github.com/PeCatch-Artifact/PeCatch-Artifact}}.

\section{Background and Related Work}

This section describes the background of this project, including Ethereum, 
the Solidity programming language, gas wastes,
and existing techniques related to ours.

\subsection{Ethereum and the Gas Mechanism}
\label{sec:eth}

As a blockchain system, Ethereum empowers programmers to write smart contracts for 
dapps~\cite{eth-1,eth-2}. Ethereum's native cryptocurrency is Ether 
(ETH). Within Ethereum, both users and smart contracts are distinct accounts with unique 
addresses for interaction. An Ethereum transaction involves a series of computations 
initiated by a message from a user account, ranging from simple Ether transfers to 
complex computations implemented by multiple smart contracts. The EVM serves as the 
execution environment for transactions, 
ensuring consensus upon transaction commitment.

Gas represents the cost of executing computations on Ethereum, 
and users must pay gas fees for 
their transactions~\cite{gas-1:url,gas-2:url}.  
When submitting a transaction, 
the user specifies both the maximum gas limit
and the priority fee she is willing to pay to incentivize miners to prioritize processing the transaction.
Certain gas units are charged
to exercise each instruction. 
If the gas usage exceeds the limit, the 
transaction is canceled, and any changes are rolled back. Gas prevents 
malicious computations from depleting Ethereum's resources. 

\subsection{Solidity Programming Language}
\label{sec:solidity}

Solidity, as the most popular programming language for writing smart contracts~\cite{solidity-top-1:url}, features language elements tailored to reflect smart contracts' semantics. The primary building block in Solidity is 
\texttt{contract} (similar to \texttt{class} in Java), encompassing fields for storing states, functions for functionalities, and events for logging. Solidity provides various types, including primitive types like integer types and \texttt{address}, and complex types like \texttt{struct} and \texttt{mapping}, enabling diverse transaction protocols. It offers a complex data store system with four distinct areas\footnote{We disregard the store area transient in this paper, 
as it was introduced very recently and is mainly used for implementing reentrancy guards.
}, each with its own set of instructions for interaction and gas-cost model.

{
\begin{figure}[t]

\begin{minipage}{\columnwidth}
\begin{center}
\footnotesize
\begin{lstlisting}[xleftmargin=.15in,language=Solidity,basicstyle=\ttfamily,morekeywords={-},morekeywords={+},keepspaces=true]
contract Warehouse {
  mapping(address => uint) private _bal;
  uint _sig;
  struct CTX {uint _sig;}
  function transfer(address from, address to, uint amount, CTX calldata data) external {
    CTX memory ctx = data; //621 gas units
    require(_sig == ctx._sig); //2142 gas units
    require(_bal[from] >= amount); //223 units 
    uint old = _bal[from]; //208 gas units
    unchecked { old -= amount;} //21 gas units
    _bal[from] = old; //214 gas units
    _bal[to] += amount; //22415 gas units
  }}
\end{lstlisting}
\mycaption{fig:example}{A smart contract example with a gas waste.}
{\textit{}}
\end{center}
\end{minipage}
\vspace{-0.2in}
\end{figure}
}

\italicparagraph{Stack:}
The EVM operates as a stack-based machine, lacking registers. The stack serves as the 
primary area for interacting with instructions, providing operands and receiving results. 
It can hold a maximum of 1024 32-byte words, with direct accesses limited to the topmost 
16 words. Most stack operations incur only two or three gas units each, making it the 
most economical store area.

\italicparagraph{Memory.}
Memory is designated for holding complex types (\eg, \texttt{struct}) that cannot reside on the stack. It functions as a byte-addressable array, with operations like \texttt{mload} for reading data from the memory and \texttt{mstore} and \texttt{mstore8} for modifying the memory. Memory is not persistent across transactions, and accessing it incurs slightly higher costs than the stack, unless memory expansion happens.
For example, both \texttt{mload}
and \texttt{mstore} consume three gas units.

\italicparagraph{Storage.}
Storage stores the persistent state variables (contract fields) of contracts across transactions. It operates as a key-value store, with keys assigned unique IDs following the declaration order of contract fields. Solidity optimizes storage usage by packing multiple consecutive contract fields into single words if they are smaller than 32 bytes. Interactions with storage, such as \texttt{sload} and \texttt{sstore}, consume significantly more gas compared to other areas.

\italicparagraph{Calldata.}
Calldata facilitates data transfer from the caller to the callee when invoking a function in a different contract. It operates as a byte-addressable array, with the first four bytes reserved for identifying the callee function, while the remaining bytes store arguments. Instructions like \texttt{calldataload} and \texttt{calldatacopy} enable reading values from the calldata. Unlike memory, the calldata area is read-only, with no instructions available for modification. Reading data from calldata incurs the same gas cost as reading data from memory.

Figure~\ref{fig:example} shows an example of Solidity code 
featuring contract \texttt{Warehouse}. \texttt{Warehouse} uses 
its \texttt{\_bal} field in line 2 to monitor the quantity of tokens held by each address. 
The \texttt{transfer()} function in line 5 transfers tokens between addresses, 
with its four parameters representing sender, receiver, 
transferred amount, and context, respectively. 

This code involves all four store areas. 
All contract fields (\eg, \texttt{\_bal} in line 2) 
are on the storage.
The first three function parameters are on the stack.
Moreover, the results of reading from a storage slot (\eg, line 9) and 
mathematical computations
are also on the stack (\eg, line 10).
The \texttt{calldata} parameter \texttt{data} is copied to the \texttt{memory}
struct \texttt{ctx} in line 6.
Gas consumption for each source-code line is also shown in the figure. 
Reading or writing a storage slot (\eg, line 9, line 11) consumes more than 200 gas units.
The write in line 12 consumes significantly more gas than line 11, because it alters 
a storage slot from zero to non-zero. 
If the \texttt{unchecked} in line 10 is removed, an underflow check will be added by the compiler, resulting in gas consumption of more than 100. 
Line 6 involves memory allocation and boundary check operations. Thus, its gas consumption exceeds 600 units. 

\subsection{Gas Wastes}
\label{sec:gas}

Ethereum calculates gas usage intricately, with various opcodes consuming differing amounts of gas. Accessing data from different storage areas incurs varied gas costs (\eg, storage access is pricier than stack access). 
Moreover, even the same opcode on the same store area can have different 
gas consumption. For example, changing a storage word 
from $0$ to $1$ consumes over 10K gas units – 
significantly more than changing it between non-zero values. Consequently, Solidity programmers may inadvertently write gas-inefficient code. Furthermore, Ethereum faces millions of dollars in daily gas fees, highlighting the crucial need to optimize gas usage for Solidity programs~\cite{gastracker}.

In this paper, we define ``gas wastes'' as code segments that unnecessarily consume excessive gas. These inefficiencies can be fixed without compromising original functionalities.
Our goal 
is to decrease gas costs by detecting and fixing gas wastes
before deploying contracts on Ethereum. 
Subsequently, after deployment, contract users will pay lower gas fees for equivalent computations, which could increase their willingness to interact with the contracts.

Gas wastes resemble performance bugs in traditional programming languages~\cite{PerfBug,perf.fse10,ORMPatterns,SmartphoneStudy,CLARITY} 
in that neither affects the correctness of the program outcome, but gas wastes differ in several ways.
First, their impacts are measured in different ways: gas wastes are measured by the monetary cost of running a contract, 
but performance bugs are measured by the time (latency, throughput) to run a program.
Second, different data store areas in Solidity manage data with varying lifespans (\eg, 
persistent data is stored in storage, and stack is used for temporary data within a function), and contracts can access these store areas directly. 
Additionally, the type of data store operations performed significantly influences the gas costs of executing a contract. 
On the other hand, a program's memory system organizes data based on access frequency. When data from a distant store area is accessed by the processor, it is automatically cached in a closer store area to improve efficiency.
%
%
Third, the EVM differs from traditional architectures (\eg, JVM) 
by having its own unique opcodes (\eg, opcodes that allow direct access 
to the persistent store area), 
while lacking certain opcodes and hardware features (\eg, registers or 
local variables that function like registers) typically found in those traditional architectures. 
As a result, some gas wastes are unique to EVM.
%
Therefore, study results of performance bugs cannot be applied to gas wastes and the smart-contract environment.

\begin{table}[t]
\centering
\footnotesize
\mycaption{tab:apps}
{Information of Selected Solidity Applications.}
{\textit{(Unique: wastes unique to Solidity.)}
}
\vspace{0.05in}
{
 \setlength{\tabcolsep}{0.8mm}{
\begin{tabular}{|l|c|c|c|c|c|c|c|c|}
\hline
{\textbf{Apps}}   &  {\textbf{Stars}}  	& {\textbf{Commits}} & {\textbf{LoC}} & {\textbf{Contracts}} & {\textbf{Wastes}}  & {\textbf{Unique}}    \\ \hline \hline
OpenZeppelin	& 22623	& 3311	& 11457	& 280 & 23 & 14 \\ \hline
Uniswap V3	    & 3725	& 1005	& 3382	& 62 & 27 & 13\\ \hline
uniswap-lib	    & 140	& 73	&  666	& 17 & 4  & 1	        \\ \hline
solmate	        & 3231	&  426	& 6840 & 36 &25  & 13 \\ \hline
Seaport	        & 1992  & 5400 & 6646 & 62 & 21 & 13 \\ \hline
\end{tabular}
}
}

\vspace{-0.15in}
\end{table}

\subsection{Related Work}
\label{sec:related}

\bolditalicparagraph{Gas Usage Optimization.}
Gas usage optimization is a crucial focus for the official Solidity language team.
In version 0.8.0, Solidity introduced automated overflow and underflow checks for
all mathematical operations. Additionally, it offered the ``unchecked'' feature, allowing
programmers to mark specific code regions to
disable the checks and save gas~\cite{unchecked}.
The Solidity compiler provides two types of gas usage optimizations~\cite{solidity-optimizations}.
Unfortunately, these optimizations are conservative and 
utilize algorithms commonly used in traditional programming language compilers. 
As a result, they cannot address most of the gas wastes 
discussed in Section~\ref{sec:study}. 
Furthermore, we have no indication that these wastes will be optimized in future releases of the Solidity compiler.

Researchers have developed multiple detection techniques to 
identify gas wastes in Solidity 
programs~\cite{brandstatter2020characterizing,nelaturu2021smart,chen2018towards,gaschecker,chen2017under,slither,synthesissmt, superoptimizer,grech2018madmax,ghaleb2022etainter,kong2022characterizing}.  
Although useful, these techniques either solely 
apply traditional compiler optimizations to Solidity, overlooking the language's 
unique features (\eg. python-solidity-optimizer), or concentrate on detecting only 
a restricted set of gas-inefficient code patterns (\eg, gas wastes within
on single basic block~\cite{superoptimizer}). 
As a result, these methods fall short of detecting the majority of gas wastes in the real world.
For example, in our evaluation, MadMax~\cite{grech2018madmax} does not detect any gas
wastes.


Previous works also offer general suggestions for saving gas, such as reducing the amount of data stored on the storage~\cite{marchesi2020design} and sacrificing contract readability~\cite{zou2019smart}.
Unlike these studies, our research is based 
on specific issues collected from contract repositories and online traces, making our findings more concrete and actionable.

Researchers have also examined gas usage when deploying smart contracts~\cite{vacca2022empirical, ajienka2020empirical}. 
However, our primary focus is on gas consumption during contract execution. 

Overall, current compiler optimizations and detection techniques are not sufficiently effective at capturing real-world gas wastes, which is 
largely due to the lack of empirical studies on gas wastes. 
Therefore, we perform such a study. 
Our findings can guide future research on resolving gas waste, 
as demonstrated by the results in Section~\ref{sec:detect}.

\bolditalicparagraph{Other Solidity Research.}
Researchers have conducted studies to comprehend various aspects
of Solidity programs, including inline assembly code and loops in them~\cite{chaliasos22astudy,bartoletti2017anempirical}, their functionalities and associated design patterns~\cite{bartoletti2017anempirical}, their processed on-chain data~\cite{pinna2019amassive}, 
and their measurements on source-code metrics~\cite{tonelli18smart}. However, none of them try to understand gas-inefficient code patterns. 
%
Researchers build many techniques to pinpoint different types of Solidity bugs~\cite{liu2018reguard, qian2020towards,xue2020cross,wang2019detecting, li20safepay,yang2021finding,chen2023tyr,wust2016ethereum, xu2020eclipsed, marcus2018low},
%
but these techniques mainly improve Solidity programs' safety and security, 
not gas efficiency.

\section{Gas Wastes in Real Solidity Programs}
\label{sec:study}

This section presents our empirical study on gas wastes collected from real-world Solidity programs, including the study's methodology, the underlying root causes of the gas wastes, and their fixing strategies.

\begin{table}[t]
\centering
\footnotesize

\mycaption{tab:root}
{Solidity-specific gas wastes categorized by store areas and reasons for not being optimized.}
{\textit{(Lack: lack of optimizations; Trade-off: trading gas usages for others;  
Issue: implementation issues in the Solidity compiler.)}}
\vspace{0.05in}
{
\setlength{\tabcolsep}{1.5mm}{
\begin{tabular}{|l|c|c|c|c|}
\hline
\multirow{2}{*}{\textbf{Store Areas}} & \multicolumn{3}{c|}{\textbf{Why Not Optimized?}} &\multirow{2}{*}{{\textbf{Total}}}       \\ \cline{2-4}
  & {{\textbf{Lack}}} & {{\textbf{Trade-off}}} & {{\textbf{Issue}}}  & \\ \hline \hline

\textbf{Stack}          & $2$  & $14$ & $8$   & $24$   \\ \hline
\textbf{Memory}         & $4$  & $0$ & $0$   & $4$   \\ \hline
\textbf{Storage}        & $20$   & $0$ & $0$   & $20$   \\ \hline 
\textbf{Calldata}       & $6$   & $0$ & $0$   & $6$   \\ \hline 
\hline
\textbf{Total} & $32$  & $14$ & $8$ & $54$ \\ \hline

\end{tabular}
}
}

\vspace{-0.2in}
\end{table}

\subsection{Methodology}
\label{sec:meth}

We gather the studied gas wastes from five 
open-source Solidity applications in Table~\ref{tab:apps}.
We select these five applications for several reasons. 
First, they are popular GitHub repositories. 
For instance, OpenZeppelin~\cite{OpenZeppelin:url} boasts 22K GitHub stars.
Second, these applications serve as fundamental components 
for numerous essential blockchain applications. Consequently, analyzing gas inefficiency patterns within them can yield a substantial impact. 
For instance, 
Uniswap-lib~\cite{uniswap-lib:url} 
is shared among all Uniswap contracts. 
Third, these applications encompass common functionalities~\cite{lo2021uniswap, sun2023demystifying, khan2022code} of Solidity programs, 
such as mathematical computations, token trading, and access control. 
Their code 
accurately reflects typical 
coding practices used by Solidity programmers.

To collect gas wastes, we employ a set of keywords to search through 
the GitHub commit logs of the applications, an established method for 
identifying real-world bugs~\cite{PLDI20Understanding,go-asplos}. 
Initially, we use ``gas’’ and ``opt’’ as keywords. 
As we inspect more gas wastes, we iteratively expand 
our keyword set to include ``store,’’ ``load,’’ and ``uncheck’’ 
to ensure comprehensive coverage. Overall, we find 371 
commits containing the five keywords. 
We then manually analyze the search results to identify commits 
fixing gas wastes and verify that the identified gas wastes can\textit{not} 
be optimized by the Solidity compiler under the ``\texttt{-{}-optimize-runs}'' option or  the ``\texttt{-{}-via-ir}'' option
using toy programs. 
In total, we collect 100 instances of gas waste, as shown in Table~\ref{tab:apps}.

To gain a comprehensive understanding of a gas waste, we primarily rely on the information in the corresponding commit. 
We thoroughly examine the modified code and its surrounding code context. 
Moreover, we also pay careful attention to the textual descriptions 
accompanying the commit and discussions among programmers 
in the related pull request and issue report.
Each gas waste is studied by at least two paper authors independently. 
All study results are thoroughly discussed in multiple rounds to solve disparities.

We employ a two-step approach to investigate the root causes of gas wastes. 
Initially, we identify gas wastes that are specific to Solidity,
by considering whether or not the same code can lead to performance issues 
in other programming languages as well. 
As shown in Table~\ref{tab:apps}, 54 out of the 100 gas wastes
are associated with Solidity's unique language features. 
The remaining 46 gas wastes
can be detected using existing compiler optimizations 
and performance bug detection techniques~\cite{Alabama,CARAMEL,XuDataStructure,BloatFSE2008,jolt,XuBloatPLDI2009,XuBloatPLDI2010,LeakChaser,Cachetor,LoopInvariant,PerfBlower,Reusable,Resurrector,PerfBug,perf.fse10,SmartphoneStudy,CLARITY,yufei-perf,he2022perfsig,liu2022automatic,chen2022adaptive,ding2020towards,yang2018how,su2019pinpointing,zhao2020auto,inline-asplos22}, 
or they can be optimized with better algorithms, 
regardless of programming languages or underlying architectures. 
For instance, six gas wastes can be patched by inlining a callee function, 
and seven gas wastes are resolved by adopting new mathematical algorithms. 
Since these gas wastes have been covered by the previously 
cited papers, we focus our analysis on wastes unique to Solidity.

Next, we categorize the 54 gas wastes specific to Solidity by addressing the following questions.
First, \textit{where is the data manipulated by the gas-inefficient code stored?}
Since different store areas involve distinct interaction instructions, leading to various 
buggy code patterns for corresponding gas wastes, organizing gas wastes according to store 
areas can guide the design of gas-waste detection methods.
Second,	\textit{why can't the Solidity compiler optimize these gas wastes?}
Answering this question helps us understand the limitations of the current toolchain and 
provides insights into potential improvements.
Third, \textit{what strategies are employed to fix these wastes?}
By answering this question, we aim to identify effective methods to optimize gas usage.

\subsection{Store Areas}

As shown in Table~\ref{tab:root}, we categorize the 54 gas wastes based on their occurrence areas.

{
 \begin{figure}[t]

 \begin{minipage}{\columnwidth}
 \begin{center}
 \footnotesize
\begin{lstlisting}[xleftmargin=.15in,language=Solidity,basicstyle=\ttfamily,morekeywords={-},morekeywords={+},keepspaces=true]
 contract ERC20 {
   uint public totalSupply;
   mapping(address => uint) public balanceOf;
   function transfer(address to, uint val) external returns (bool) {
     balanceOf[msg.sender] -= val;
-    balanceOf[to] += val;
+    unchecked { balanceOf[to] += val; }
     return true;
   }
   function _mint(address to, uint256 val) internal {
     totalSupply += val;
-    balanceOf[to] += val;
+    unchecked { balanceOf[to] += val;}
   }}
\end{lstlisting}
 \mycaption{fig:unchecked}{A gas waste fixed by \texttt{unchecked} in OpenZeppelin.}
{
}
\end{center}
\end{minipage}
\vspace{-0.2in}
\end{figure}
}

\subsubsection{Stack}
\label{sec:stack-cause}
The EVM is a stack-based machine, without any register. 
Thus, the stack is the most frequently accessed store area. For example, 
all mathematical operations in Solidity take operands from
and output results back to the stack. As a result, 
the ``stack’’ category has the most gas wastes. 

\italicparagraph{Not using unchecked when possible.}
In order to achieve good security guarantees, Solidity provides various compiler checks.
For example, the Solidity compiler 
automatically adds overflow and underflow checks to stack variables before each mathematical computation to prevent overflow-based attacks.
These checks consume additional gas. 
To avoid this gas consumption, Solidity allows programmers to mark operations 
with the \texttt{unchecked} keyword to disable the checks (\eg, line 10 in Figure~\ref{fig:example}). 
One execution of an \texttt{unchecked} operation saves 191 gas units.
As using \texttt{unchecked} relaxes the security guarantee, it is intended to be used only when programmers are certain that no overflow or underflow could occur at the marked operation. 
Please note that Solidity's \texttt{unchecked} feature differs 
from the \texttt{checked} and \texttt{unchecked} features in C\#. 
In C\#, overflow checks are not performed by default on most integral-type arithmetic computations, and the \texttt{unchecked} keyword is primarily used by 
programmers to indicate that overflows in the enclosed computations are inconsequential.

{
\begin{figure}[t]

\begin{minipage}{\columnwidth}
\begin{center}
 \footnotesize
\begin{lstlisting}[xleftmargin=.15in,language=Solidity,basicstyle=\ttfamily,morekeywords={-},morekeywords={+},keepspaces=true]
 abstract contract ERC4626 {
   mapping(address=>mapping(address=>uint)) allow;
   function redeem(uint shares, address to, address from) public {
     uint a = allow[from][msg.sender];
-    if (msg.sender!=from && a!=type(uint).max) {
-      allow[from][msg.sender] = a - shares;
-    }
+    if (msg.sender != from) {
+      if (a != type(uint).max) {
+        allow[from][msg.sender] = a - shares;
+      }}
   }}
\end{lstlisting}
\mycaption{fig:andinif}{A gas waste caused by using \&\& in an \texttt{if} condition.}
{}
\end{center}
\end{minipage}
\vspace{-0.2in}
\end{figure}
}

12 stack-based gas wastes are about not enabling the \texttt{unchecked} feature when possible.
In four of the gas wastes, the program semantic ensures that the result of an addition
is smaller than another number in the same type or another addition result that has been checked for overflow (and similarly, a subtraction result is bigger than another number 
or the result has already been checked for underflow). In such cases, the operation 
under examination does not need to be checked and can be marked as \texttt{unchecked} but was not, resulting in gas wastes.
An example is illustrated in Figure~\ref{fig:unchecked}.
The \texttt{totalSupply} field (line 2) holds the total number of tokens among all addresses.
\texttt{balanceOf} in line 3 is a map tracking the number of tokens for each address. 
The addition in line 6 cannot overflow since the added value is the number of tokens belonging to another address, and adding it is still smaller than or equal to \texttt{totalSupply}. 
Furthermore, the addition in line 12 does not overflow, 
since it is smaller than the addition result in line 11, 
and line 12 can only be executed when line 11 passes the overflow check.

For five other gas wastes, the programmers understand 
the workload and are confident that certain computations 
will not overflow, so they mark those operations as \texttt{unchecked}. 
Identifying these inefficiencies relies on the 
programmers' in-depth knowledge of the contracts, making it difficult to use static analysis to detect similar cases.

For the last three gas wastes, there are control flow checks to guarantee when 
the program reaches a mathematical operation, the values of the operands 
satisfy certain conditions, and thus the operation cannot overflow or underflow (\eg, line 10
in Figure~\ref{fig:example}). 



\stepcounter{insight}
\boldunderparagraph{Insight \arabic{insight}:}
{\it{
Programmers miss many opportunities of leveraging language features for trading security for lower gas.
}}

\stepcounter{suggestion}
\boldunderparagraph{Suggestion \arabic{suggestion}:}
{\it{
Static analysis tools could be developed to help programmers leverage program features at the development phase or the testing phase.
}} 

Although Zou et al.~\cite{zou2019smart} also report the need for powerful tools for 
Solidity, their focus is primarily on security, which differs from our perspective.


\italicparagraph{Using \texttt{\&\&} in if conditions.}
The EVM implements many opcodes solely using the stack. 
For example, when evaluating the ``\texttt{\&\&}'' operator 
in an \texttt{if} statement's condition, 
it first pushes the evaluation result (denoted as ``\texttt{c1}’’) 
of the left-hand condition (\eg, ``\texttt{msg.sender!=from}'' 
in Figure~\ref{fig:andinif}) 
onto the stack and then copies this 
result (denoted as ``\texttt{c2}’’) onto the stack again. 
Subsequently, it consumes \texttt{c2} by checking whether it is true. 
If it is, \texttt{c1} is popped before evaluating 
the right-hand condition. 
If \texttt{c2} is false, \texttt{c1} is checked again to decide 
whether to execute the \texttt{if} or  \texttt{else} branch. 
The copy, pop, and inspection of \texttt{c1} are all unnecessary 
and consume gas. This gas consumption can be avoided 
by splitting an \texttt{if} statement with  \texttt{\&\&} into two nested 
\texttt{if} statements. This saves 20 gas units when the 
first logical expression evaluates to false. 
Four instances of gas waste result from using  \texttt{\&\&} in an \texttt{if} statement's condition.
One simplified instance and its patch are shown in Figure~\ref{fig:andinif}.

{
\begin{figure}[t]

\begin{minipage}{\columnwidth}
\begin{center}
\footnotesize
\begin{lstlisting}[xleftmargin=.15in,language=Solidity,basicstyle=\ttfamily,morekeywords={-},morekeywords={+},keepspaces=true]
 library SafeMath {
-  function mul(uint a, uint b) public returns (uint) {
-    uint c = a * b;
+  function mul(uint a, uint b) public returns (uint c) {
+    c = a * b;
     return c;
   }}
\end{lstlisting}
\mycaption{fig:return-local}{A gas waste due to returning a local variable.}
{
}
\end{center}
\end{minipage}
\end{figure}
}

\italicparagraph{Returning local variables in a function.}
Solidity allows developers to declare the 
return value (denoted as ``\texttt{retD}'') 
of a function in the
function’s declaration. Programming in this way saves gas 
compared to returning a local variable (denoted as ``\texttt{retLocal}''),
which was also discussed in a previous paper~\cite{di2022profiling}.
For example, line 6 in Figure~\ref{fig:return-local} returns a local variable \texttt{c} for the buggy code
and it consumes more gas than declaring \texttt{c} in the function’s declaration, illustrated
by the patch in line 4. 
Four gas wastes are due to using ``\texttt{return retLocal}''.
The amount of saved gas for fixing one such waste 
depends on the type of the return. 
For example, 19 gas units are saved for 
an integer and 16 are saved for a string.


{
\begin{figure}[t]

\begin{minipage}{\columnwidth}
\begin{center}
\footnotesize
\begin{lstlisting}[xleftmargin=.15in,language=Solidity,basicstyle=\ttfamily,morekeywords={-},morekeywords={+},keepspaces=true]
 abstract contract ERC1155 {
   function safeBatchTransferFrom(uint[] memory ids, uint[] memory amounts) public {
     uint len = ids.length; 
+    uint id;
+    uint amount;
     for (uint i = 0; i < len; i++) {
-      uint id = ids[i];
-      uint amount = amounts[i];
+      id = ids[i];
+      amount = amounts[i];
     }
   }}
\end{lstlisting}
\mycaption{fig:loop-alloc}{A gas waste due to allocating stack variables in a loop.}
{}

\end{center}
\end{minipage}
\end{figure}
}

\italicparagraph{Allocation in a loop.}
As discussed in Section~\ref{sec:solidity}, 
Solidity restricts direct access to only the top 16 words on the stack. 
To access other stack slots, Solidity must first remove some elements from the stack. 
However, when there are many local variables used 
in a nested manner, the Solidity compiler
may not find a way to generate code, resulting in a ``stack too deep'' compiler error. 
To prevent such errors, Solidity identifies unused local variables and deallocates them earlier. 
For instance, Solidity removes a local variable from the stack at the closing 
curly bracket of the code block declaring the variable, 
instead of deallocating the variable
at the end of the function, like C/C++. 
Unfortunately, if the code block is within a loop, 
continuously creating and deallocating a local 
variable on the stack can lead to gas wastage. 
This situation is different from CS1 in~\cite{di2022profiling}, where CS1 involves assigning 
a new value to a variable in a loop without using the new value in the loop.
This root cause leads to two gas wastes. 
Moving the allocation of a local variable outside the loop saves 
15 gas units per iteration. Figure~\ref{fig:loop-alloc} shows one gas waste in this category
where two local variables, \texttt{id} and \texttt{amount}, are declared inside a loop 
and then popped out at the end of each loop iteration. By relocating the 
declaration site outside the loop, the allocation and deallocation of these 
variables occur just once, resulting in a reduction in gas usage.


\italicparagraph{Other causes.}
Two gas wastes do not belong to the above cases. 
One is related to bit shifting.
In Solidity, a \texttt{bytes32} number is 32 bytes long, and an \texttt{address}
is 20 bytes. This gas-inefficient code left-shifts a \texttt{bytes32} number
by 12 bytes and extracts the most-significant 20 bytes as an address, 
consuming 31 more gas units than necessary, since
the same result can be gained by extracting the least-significant 20 bytes directly.
The final gas waste is about replacing ``\texttt{a!=b}'' with ``\texttt{a<b}'', 
since ``\texttt{!=}'' is implemented with two opcodes and thus consumes more gas.
The operand ``\texttt{b}'' in the original program is the largest \texttt{uint256} number. Thus, ``\texttt{a>b}'' can never happen, allowing the statement to be reduced to ``\texttt{a<b}''.

\stepcounter{insight}
\boldunderparagraph{Insight \arabic{insight}:}
{\it{
As EVM does not have registers, the implementation of many traditional types of operations in Solidity implicitly uses stack. Programmers are easy to overlook such usage, causing gas wastes.
}}

\subsubsection{Memory}
\label{sec:memory-cause}

{
\begin{figure}[t]

\begin{minipage}{\columnwidth}
\begin{center}
\footnotesize
\begin{lstlisting}[xleftmargin=.15in,language=Solidity,basicstyle=\ttfamily,morekeywords={-},morekeywords={+},keepspaces=true]
 contract ConsiderationPure is ConsiderationBase {
   function _applyRslv(Orders[] memory orders) public {
-    for (uint i = 0; i < orders.length; ++i) {
+    uint arraySize = orders.length;
+    for (uint i = 0; i < arraySize; ++i) {
     }
   }}
\end{lstlisting}
\mycaption{fig:loop-invar}{A gas waste due to reading a loop-invariant \texttt{memory} object field in a loop.}
{
}
\end{center}
\end{minipage}
\end{figure}
}
Four gas wastes are in the \texttt{memory} category. 
Three wastes are caused by reading 
a loop-invariant object field inside a loop, and the object is on  the memory.
For example, \texttt{orders.length} in line 3 of Figure~\ref{fig:loop-invar} 
is a field of a \texttt{memory} object, and it is read in each loop iteration without being modified within the loop.
Wastes in this type are fixed by caching the field in a stack variable and reading 
the stack variable in the loop instead of reading the field. 
Caching a memory object field saves around 3 gas units for each loop iteration. 
%
This optimization is specific to  
Solidity, as the amount of gas spent is dependent on the store area.
The remaining gas waste is due to the use of \texttt{mstore} 
that modifies 32 bytes when translating all letters in a string. 
For every letter, the gas-inefficient code left-shifts 
the translated letter to the right-most byte location of a 32-byte word before 
using \texttt{mstore} to write the 32-byte word to the memory. 
To save gas, using \texttt{mstore8} to update one byte at a time 
can eliminate the left-shift operation.

\stepcounter{insight}
\boldunderparagraph{Insight \arabic{insight}:}
{\it{
There are fewer memory-related gas wastes than other types likely because memory and its relationship to stack are similar to traditional heap and thus more familiar to programmers than other data-store types.
}}

\subsubsection{Storage}
\label{sec:storage-cause}

Among the four store areas, storage incurs the highest cost. 
Consequently, gas wastes resulting from the improper utilization of data on the storage are more likely to be perceived and addressed. 
This category comprises 20 gas wastes, ranking the second among the four store areas.

\italicparagraph{Repetitive reads to storage.}
14 gas wastes stem from repetitive access to the same storage slot.
Among them, 12 involve accessing a contract field, while the remaining two involve accessing 
an input parameter stored on the storage. 
Instruction \texttt{sload} 
is required to access storage data, and each execution of it consumes more than $100$ gas units. 
Copying the data to either the stack or the memory and then accessing the copied version
significantly reduce the gas usage, since accessing the stack or the memory once only consumes two to three gas units. 
This situation is different from CS10 in \cite{di2022profiling},
as CS10 is about minimizing the amount of data saved on the storage.
Figure~\ref{fig:example} shows one such example. \texttt{\_bal[from]} is read with an \texttt{sload} 
for the first time in line 8, and it is read again
by another \texttt{sload} 
in line 9. 
A patch can store \texttt{\_bal[from]} in a stack variable,
replacing the second \texttt{sload} 
with a read on the stack variable and saving 217 gas units per execution.


\italicparagraph{Read-after-write to storage.}
Three gas wastes are caused by writing a stack value to a contract field and 
subsequently 
reading the same field without any modifications in between. For 
example, in Figure~\ref{fig:sstore-sload}, \texttt{owner} is a contract field that is assigned the value of the input parameter 
\texttt{newOwner} in line 4 and read in line 5. As shown in the 
figure, the read operation on the contract field \texttt{owner} in line 5
can be replaced with reading the input parameter \texttt{newOwner}, saving 103 gas units. 


{
 \begin{figure}[t]

 \begin{minipage}{\columnwidth}
 \begin{center}
 \footnotesize
\begin{lstlisting}[xleftmargin=.15in,language=Solidity,basicstyle=\ttfamily,morekeywords={-},morekeywords={+},keepspaces=true]
 abstract contract Auth {
   address public owner;
   function setOwner(address newOwner) public {
     owner = newOwner;
-    emit OwnerUpdated(owner);
+    emit OwnerUpdated(newOwner);
   }}
\end{lstlisting}
 \mycaption{fig:sstore-sload}{A gas waste due to read-after-write to storage.}
{
}
\end{center}
\end{minipage}
\vspace{-0.2in}
\end{figure}
}

{
\begin{figure}[t]

\begin{minipage}{\columnwidth}
\begin{center}
\footnotesize
\begin{lstlisting}[xleftmargin=.15in,language=Solidity,basicstyle=\ttfamily,morekeywords={-},morekeywords={+},keepspaces=true]
 contract ReentrancyGuard {
-  bool private reentrancyLock = false;
+  uint256 private reentrancyLock = 1;
   modifier nonReentrant() {
-    require(!reentrancyLock);
-    reentrancyLock = true;
+    require(reentrancyGuardFree == 1);
+    reentrancyLock = 2;
     _;
-    reentrancyLock = false;
+    reentrancyLock = 1;
   }}
\end{lstlisting}
\mycaption{fig:bool}{A gas waste due to using a Boolean contract field.}
{
}
\end{center}
\end{minipage}
\end{figure}
}

\stepcounter{insight}
\boldunderparagraph{Insight \arabic{insight}:}
{\it{
Specific patterns of reading and writing to storage could always lead to extra gas.
}}

Although Sorbo et al.~\cite{di2022profiling} also report the gas wastes
due to consecutively writing a storage variable,
their code pattern requires the write is inside a loop, while we do not 
have such a requirement.

\stepcounter{suggestion}
\boldunderparagraph{Suggestion \arabic{suggestion}:}
{\it{
Static analysis of storage reads and writes could capture gas wastes and automatically fix them. 
}}

\italicparagraph{0-to-1 gas charge.}
Changing a zero to a non-zero value on storage 
consumes 100X more gas than modifying non-zero values, 
because the size of a zero value on blockchain is also zero, 
but a piece of non-zero data occupies 256 bits. 
While changing from zero to non-zero is unavoidable 
for regular values, it could be avoided for Booleans.
Changing a Boolean storage value from false to true implicitly changes the storage slot from 0 to 1. If \texttt{uint256} is used 
and 1 and 2 are used to represent false and true, one could avoid this high gas consumption 
and save 22242 gas units for one execution.
For example, as shown by Figure~\ref{fig:bool}, \texttt{reentrancyLock} is a Boolean contract field in the buggy version, 
and line 6 updates it from 0 to 1, resulting in significant gas consumption. 
The patch modifies \texttt{reentrancyLock} to an integer, 
replacing true and false with 2 and 1, respectively, in order to reduce gas usage.

Three gas wastes are caused by this Boolean usage and fixed with the above strategy.
Importantly, it is not advisable to convert all Boolean contract fields into \texttt{uint256}. 
When a Boolean field is packed with other fields in the same storage slot, all of them only consume one 256-bit storage slot.
Changing the Boolean to \texttt{uint256} increases storage size, potentially incurring
more cost when deploying the contract.

{
\begin{figure}[t]

\begin{minipage}{\columnwidth}
\begin{center}
\footnotesize
\begin{lstlisting}[xleftmargin=.15in,language=Solidity,basicstyle=\ttfamily,morekeywords={-},morekeywords={+},keepspaces=true]
 abstract contract ERC1155 {
   mapping(address => mapping(uint256 => uint256)) public balanceOf;
   function safeBatchTransferFrom(
     address from, address to,
-    uint256[] memory ids,
-    uint256[] memory amounts,
+    uint256[] calldata ids,
+    uint256[] calldata amounts,
   ) external {
     uint256 idsLength = ids.length;
     uint256 id;
     uint256 amount;
     for (uint256 i = 0; i < idsLength; ++i) {
       id = ids[i];
       amount = amounts[i];
       balanceOf[from][id] -= amount;
       balanceOf[to][id] += amount;
     }}}
\end{lstlisting}
\mycaption{fig:calldata}{A function whose parameters can be from \texttt{memory}
to \texttt{calldata}.}
{
}
\end{center}
\end{minipage}
\end{figure}
}

\stepcounter{insight}
\boldunderparagraph{Insight \arabic{insight}:}
{\it{
The gas-cost model can be tricky, requiring careful consideration to find the best way to write code.
}}

\subsubsection{Calldata}
\label{sec:calldata-cause}

Six gas wastes are caused by incorrectly labeling function arguments or returns as 
\texttt{memory} instead of \texttt{calldata}. 
Among these, four gas wastes arise when labeling 
a read-only argument of an external function as \texttt{memory}. 
External functions can 
only be invoked from functions in a different contract, 
and the argument resides in the calldata at the 
beginning of the invocation. Using \texttt{memory} results in unnecessary copying of the 
argument from calldata to memory when entering the function, which wastes gas, since the 
argument can be directly read from calldata. For example, 
changing an array parameter with 100 elements from \texttt{memory}
to \texttt{calldata} saves 23182 gas units.
Figure~\ref{fig:calldata} illustrates a gas waste caused by this issue. 
The parameters \texttt{ids} and \texttt{amounts} are only read within 
the external function \texttt{safeBatchTransferFrom}, so that their labels can be changed 
from \texttt{memory} to \texttt{calldata} to optimize gas usage.

Similarly, if an internal function is only invoked with a real parameter from the calldata 
for an argument, and the function does not modify the argument, then the argument can be labeled 
as \texttt{calldata}. Moreover, if a function returns a piece of data from the calldata, and that 
return is read-only for its callers, the return value can be labeled as 
\texttt{calldata}. Failing to label an internal function's argument and a function's return as 
\texttt{calldata} causes two gas wastes, respectively.

\stepcounter{insight}
\boldunderparagraph{Insight \arabic{insight}:}
{\it{
The failure of using \texttt{calldata} causes gas wastes, likely as a result of programmers' unfamiliarity with the new \texttt{calldata} type in Solidity but not in traditional languages.
}}

\stepcounter{suggestion}
\boldunderparagraph{Suggestion \arabic{suggestion}:}
{\it{
Static analysis of function arguments and returns can avoid calldata-related gas wastes. 
}} 


GasSaver~\cite{gassaver} detects function parameters that can be changed from \texttt{memory} to \texttt{calldata}, 
thus implementing this suggestion. 
However, as shown by the evaluation results in Section~\ref{sec:detect}, GasSaver does not cover all gas wastes in this category, 
highlighting the necessity of our study.

\subsection{Why Not Optimized}
\label{sec:not-optimized}

As shown in Table~\ref{tab:root}, 
we categorize the reasons for the escape of the gas 
wastes from the optimization process into three groups. 
We tackle wastes in different groups with distinct policies to complement 
the optimization process in Section~\ref{sec:detect}.

\italicparagraph{Lack of compiler optimizations.}
Approximately 60\% of the gas wastes are 
attributed to the Solidity compiler's lack of specific optimizations,
including all wastes on the memory, the storage, and the calldata. 
For instance, the compiler does not check if two \texttt{sload} operations 
access the same storage slot and if the slot is written to between the two \texttt{sload}s. 
As a result, it misses opportunities to optimize 
repeated reads on the same storage location.
Additionally, there are two instances of stack wastes in this category: 
the compiler fails to optimize left-shifting and extracting the most significant bytes 
to obtain the least significant bytes, 
and it does not replace ``\texttt{a!=b}'' with ``\texttt{a<b}'',
when \texttt{b} is the largest \texttt{uint256} number.

\italicparagraph{Trading gas usages for other properties.}
The compiler deliberately generates gas-inefficient opcodes for 14 stack wastes 
to trade off runtime gas consumption for 
improved contract reliability. Those wastes
are due to the unique language features of Solidity. 
12 of them occur because the compiler 
conservatively adds overflow or underflow checks before mathematical 
operations that cannot overflow or underflow. 
The remaining two result from the repeated allocation and deallocation of stack variables within a loop, rather than freeing all stack variables only once at the end of the function, to prevent the ``stack too deep'' error.

\italicparagraph{Implementation Issues.}
Eight stack wastes 
result from implementation issues in the Solidity compiler. Four wastes 
where  \texttt{\&\&} is used in an \texttt{if}'s condition 
are due to the compiler evaluating \texttt{if} conditions 
in a standalone function and determining whether to 
execute the \texttt{if} or \texttt{else} branch in another function. 
This approach prevents the common practice of using the result of
the first logical expression to decide \textit{both} whether
to evaluate the second logical expression and which branch to take. 
In the four wastes due to returning \texttt{retLocal}, 
the compiler copies \texttt{retLocal} to the stack slot representing the return value instead of returning \texttt{retLocal} directly.

\stepcounter{insight}
\boldunderparagraph{Insight \arabic{insight}:}
{\it{
Stack operations are complex, and even Solidity 
compiler developers can make mistakes on them when making design trade-offs and implementing the compiler.
}}

\subsection{Fixing Strategies}

We divide the gas wastes' fixing strategies into 
four groups.

\italicparagraph{Changing store areas.}
A total of 26 gas wastes are fixed by making changes to the store area
for the gas-inefficient code' manipulated data. 
Out of these, 17 gas wastes are resolved by replacing the accesses to storage data with accesses to stack or memory data. 
%
All six gas wastes that resulted from failing to label a \texttt{memory} argument 
as \texttt{calldata} are fixed by modifying the respective arguments' labels. 
Lastly, the three gas wastes arising from accessing loop-invariant memory 
data are resolved by moving the data to the stack.

\italicparagraph{Changing stack operations.}
Nine gas wastes are patched by changing the way of implementing semantics with stack operations. 
They are the gas wastes caused by \texttt{\&\&} in \texttt{if} conditions, returning local variables ``\texttt{retLocal}'', and stack-variable allocation in loops, as explained in Section~\ref{sec:stack-cause}. 
Their fixes are using nested \texttt{if} statements, changing function returns to ``\texttt{retD}'', and moving allocation outside loops.

\italicparagraph{Eliminating computation.}
15 gas wastes are resolved by eliminating computation. 
Among these, 12 gas wastes 
are fixed by adding the \texttt{unchecked} tag to computation for avoiding compiler overflow/underflow checks (\eg, Figure~\ref{fig:unchecked}). 
For the remaining three gas wastes,  
one is fixed by avoiding a left-shift operation and directly extracting the corresponding bits, 
one is fixed by replacing the two-opcode operation ``\texttt{a!=b}’’ with the single-opcode operation ``\texttt{a<b}'', and
the last waste is fixed by using \texttt{mstore8} to replace \texttt{mstore} to avoid
left-shift operations when modifying each letter in a string.

\italicparagraph{Others.}
Four gas wastes fall outside the scope of the previous categories, including
three wastes fixed by replacing Boolean contract fields with \texttt{uint256} contract fields, 
and the final 
waste fixed by implementing a smart algorithm to compute an \texttt{if} condition that involved the ``\texttt{\&\&}’’ operator.

\stepcounter{insight}
\boldunderparagraph{Insight \arabic{insight}:}
{\it{
Most of the gas wastes caused by misuse of memory, storage, and calldata are resolved by modifying store areas. 
Differently, stack-related gas wastes are fixed by changing or eliminating stack-related operations. 
This indicates that non-stack-related  wastes could more easily be automatically fixed. 
}}
\section{Gas Wastes in On-Chain Traces}
\label{sec:chain}

Section~\ref{sec:study} examines gas wastes fixed by programmers 
in real open-source Solidity projects. 
However, some wastes may have gone unnoticed by developers and, 
consequently, remain unresolved in Solidity programs. 
To identify and investigate these overlooked wastes, 
we analyze Ethereum execution traces in this section. 

\subsection{Methodology}
We gather on-chain Ethereum transaction traces using the debug API of Geth, an Ethereum execution client~\cite{geth:url}. 
The API returns a sequence of opcodes executed 
by each transaction. For each opcode, the API provides details 
 such as the gas consumed, the contents of the stack and 
calldata before executing the opcode, and the contents of the 
memory and storage after executing the opcode. We also infer operand values from the values at the top of the stack.

Overall, we analyze all 10,629,589 successful transactions executed on Ethereum 
between Feb 1st, 2024, and Feb 10th, 2024. 
These transactions span 71,234 blocks, consume 768.8 billion gas units, and
incur more than \$160 million in gas fees\footnote{
An Ether is \$3,659 and a gas unit is 56.89 Gwei in our measurement.
}. 
As the number of blocks and executed transactions 
remain relatively stable across days, ten days' data is ample for our study, especially given the large number of transactions.

We use a methodology similar to the one in Section~\ref{sec:study} to identify and
categorize gas wastes.
Specifically, we manually examine the opcodes executed in various store 
areas to identify gas-inefficient opcode sequences (gas wastes) 
and analyze why the compiler cannot optimize them. 
We particularly focus on frequently executed sequences~\cite{borgelt2012frequent} 
and sequences whose execution does not generate any side effects. 
The former offers large optimization rewards, 
while the latter indicates optimization 
opportunities~\cite{ldoctor}. 
We implement Python scripts to identify these two types of sequences, 
each with fewer than 20 opcodes. We then manually examine the top 
100 sequences based on their frequencies observed in the traces.
Since the gas wates studied in this section have not been 
addressed by programmers, we do not consider fix strategies here.


\subsection{Runtime Gas Wastes}

Based on the collected transaction traces, 
storage opcodes account for the majority of gas expenditure at 64.4\%. 
Although stack operations are individually inexpensive, 
their high frequency causes stack opcodes to represent 34.1\% of gas usage. 
Gas consumption for both memory and calldata is minimal, 
contributing less than 1\% of the total gas usage. Given these results, 
we focus our analysis on storage and stack opcodes and omit the analysis of memory and calldata.

\subsubsection{Storage}
The opcodes \texttt{sstore} and \texttt{sload} are used to write and read a 
storage slot, respectively. They rank first and second in gas consumption among 
all opcodes, accounting for 40.67\% and 23.72\% of total gas consumption, 
respectively. 
A significantly higher amount of gas is charged by Ethereum 
for a \texttt{sstore} when it transitions 
a storage slot from zero to a non-zero value.
We observe that 19.96\% of gas is spent on those 
initialization operations. 
Similar to a suggestion given by \cite{marchesi2020design}, 
we recommend
programmers carefully design
contract logic to reuse storage space 
or move some contract states off-chain to reduce storage usage.

\italicparagraph{Skippable \texttt{sload}.}
We identified a gas waste involving an \texttt{sload} whose result is not used 
and is subsequently popped from the stack. 
Thus, the \texttt{sload} can be skipped.
This waste occurs when resizing a dynamic array. If the array length is less than 32 bytes, both the length and the array data are stored in one storage slot; otherwise, they are in separate slots.
The compiler reads the slot with the length using an 
\texttt{sload}, extracts the length, and leaves it on the stack. 
Later, it reads the slot again with another \texttt{sload} for the 
array data and compares the length on the stack with 32 to 
decide if the second \texttt{sload}’s result contains the data. 
If the length is 32 bytes or more, the result is discarded. 
Clearly, the compiler should generate the opcode for the second 
\texttt{sload} after the opcodes comparing the array length with 
32. Thus compiler implementation issue causes the waste.

\stepcounter{insight}
\boldunderparagraph{Insight \arabic{insight}:}
{\it{
Implementation issues in the Solidity compiler can cause unnecessary storage operations, leading to wasted gas.
}}

\subsubsection{Stack}
The EVM stores all operands and computation results 
on the stack, causing the top 10 most frequently used opcodes 
being stack opcodes. 
We further observe that there are long sequences of stack opcodes. 
The longest one contains 350,035 stack opcodes.
Additionally, 15.68\% of stack opcodes are part of a sequence with 
100 stack opcodes or more. 
Since these sequences are computations independent of 
contract states, they have the potential to be replaced with off-chain 
computations to save gas.

\stepcounter{insight}
\boldunderparagraph{Insight \arabic{insight}:}
{\it{
Lengthy stack-only computations are prevalent, suggesting opportunities of contract refactoring to relocate these computations off-chain for reduced gas usage.
}}

In addition to the wastes discussed in Section~\ref{sec:stack-cause}, 
we identify 13 more wastes by analyzing stack opcode sequences.

\italicparagraph{Runtime Constant Computation.}
Nine wastes involve performing mathematical computations on 
constants at runtime instead of calculating the results during 
compilation and using the precomputed results at runtime. One 
sequence casts the number on top of the stack to an address. 
Since an address is 20 bytes long, the sequence pushes 
\texttt{0x1}, \texttt{0x1}, and \texttt{0xa0} onto the stack, 
uses \texttt{shl} and \texttt{sub} to compute $2^{161}-1$, and 
employs $2^{161}-1$ to extract the least significant 20 bytes 
from the top value on the stack. This sequence is intentionally 
generated by the compiler because $2^{161}-1$ occupies more 
space than the sequence itself. Using the sequence rather than 
the constant reduces the gas cost when deploying contracts.
Additionally, contract programmers can set the value 
of ``\texttt{-{}-optimize-runs}'' when compiling a contract to 
balance runtime gas cost against deployment gas cost. If this 
parameter is set to a value larger than 200, indicating the 
contract is expected to execute more than 200 times, 
$2^{161}-1$ will be used in the compiled opcodes. However, we 
find contracts using the sequence execute one million times, 
indicating that programmers do not always make the optimal 
configuration.

For eight wastes, the compiler pushes 
a constant (\eg, zero) onto the stack and 
then performs operations (\eg, right-shifting) on the constant. 
Those operations can be skipped without altering the program semantics. 
These gas-inefficient sequences can be removed by 
making modifications to the compiler’s implementation. Thus, we consider these wastes to be due to issues in the compiler's implementation.

\italicparagraph{Used Stack Variables.} The remaining four wastes 
arise from generating a value on the stack using \texttt{push} or \texttt{dup}.
However, these values are not utilized by subsequent 
opcodes and are eventually popped out from the stack. 
This situation is akin to 
duplicating \texttt{c1} to 
have \texttt{c2} on the stack, only to later pop 
\texttt{c1} from the stack in scenarios where 
\texttt{\&\&} is used in an \texttt{if} statement's 
condition. Similarly, we consider the four wastes to be caused by compiler implementation
errors.

\stepcounter{insight}
\boldunderparagraph{Insight \arabic{insight}:}
{\it{
Stack wastes due to compiler issues widely exist,
and many of them may not be noticed by the programmers. 
}}

\stepcounter{suggestion}
\boldunderparagraph{Suggestion \arabic{suggestion}:}
{\it{
Efforts are needed to identify implementation issues in the Solidity compiler, particularly those affecting opcode generation for stack manipulation, to prevent further gas wastage.
}}

Zou et al.~\cite{zou2019smart} also highlight the importance of identifying and 
addressing issues in the Solidity compiler, but their focus is on security 
vulnerabilities caused by the compiler, rather than gas wastes.

\section{Gas Waste Resolving}
\label{sec:detect}

We employ different strategies to address the gas wastes 
in different categories. For wastes caused by a lack of optimizations or 
by trading runtime gas usage for other properties, we build a static detection tool suite called \textit{PeCatch} to detect similar wastes in contract source code. 
Contract programmers can refer to the fixing strategies suggested in this section 
to patch the detected wastes 
before deploying their contracts. 
While we acknowledge that automatically fixing detected wastes would be beneficial, developing such a technique is outside the scope of this paper. 
For wastes caused by compiler implementation errors, 
we identify the errors in the compiler's source code 
and report them to the Solidity team. Fixing these errors in the compiler will automatically benefit all contract programmers.

\subsection{PeCatch: Detecting Gas Wastes in Source Code}
\label{sec:pecatch}

%
PeCatch comprises six distinct checkers, each crafted based on a gas-waste pattern in Section~\ref{sec:study}. 
These checkers detect gas wastes across all four store areas.

\subsubsection{Stack}
There are two checkers in this category.
The first identifies mathematical computations 
free from overflow or underflow issues (\textit{Uncheck}), 
and thus can be labeled as \texttt{unchecked}. 
%
%
%
The high-level idea is to identify 
a pair of integers ($a$, $b$), 
where it can be guaranteed (through static analysis) that $a$ is always greater than or equal to $b$ throughout the whole program. 
If $a$ ($b$) has been checked for overflow (underflow), 
then an addition (subtraction) to $b$ ($a$) 
cannot overflow (underflow) and can be marked as \texttt{unchecked}.

%
%
%

We analyze the entire contract when both $a$ and $b$ are contract fields,
and one single function when both $a$ and $b$ are local variables of the function. 
We take three steps for each scenario: identifying candidate pairs, filtering out
pairs where $a \ge b$ is not ensured, 
and pinpointing computations that can be marked as \texttt{unchecked}. 

We consider $a$ and $b$ as a candidate pair 
if they both belong to the same integer type. 
Additionally, we permit $b$ to be a collection (\eg, \texttt{mapping}) and represent 
the value of $b$ as the sum of all its elements,
as if we can safely infer that the whole collection of $b$ is no larger than $a$ and $a$ cannot overflow,
then any element of $b$ cannot overflow.
We then filter out candidates where the condition $a \geq b$ is not guaranteed.
When dealing with contract fields, this filtering begins by examining 
the initial values of $a$ and $b$ at the contract’s constructor. 
For local variables, the filtering 
inspects the initial values of $a$ and $b$
inside the function.
We only consider $a$-$b$ pairs where both are initialized to constant values and $a$'s initial constant value is no smaller than $b$'s.
Next, we examine each modification made to $a$ or $b$ within the contract (or within the function).
If $k$ is added to $b$, we verify whether $k$ is added 
to $a$ in the same function and whether $k$'s value stays the same across the two additions.
This involves inspecting whether such an addition operation exists and 
whether it either precedes and dominates 
the addition on $b$ 
or succeeds and post-dominates the addition on $b$.
For example, \texttt{val} is added to an element of \texttt{balanceOf} 
in line 12 in Figure~\ref{fig:unchecked},
which is preceded and dominated by the addition of \texttt{val} 
to \texttt{totalSupply} in line 11. 
Similarly, if $c$ is subtracted from $a$, we check whether 
$c$ is subtracted from $b$
in the same function. 
In the case where $b$ is a collection, if $c$ is subtracted from one of its elements
and added to another element, the relationship between $a$ and the sum of elements in $b$
remains unchanged (\eg, lines 5 -- 6 in Figure~\ref{fig:unchecked}).
After the filtering process, we 
identify all additions to $b$ (\eg, lines 6 and 12 in Figure~\ref{fig:unchecked})  
and all subtractions from $a$ as non-overflowable computations.


Moreover, the checker detects two extra cases with control-flow information. 
First, if a value is subtracted from a variable after checking the variable is not smaller than
the value (through an \texttt{if} check), 
the subtraction 
cannot underflow (\eg, line 10 in Figure~\ref{fig:example}).
Second, for a loop whose iteration number is less than a loop-invariant 
value,
the operation adding one to the loop-index variable
cannot overflow. 

The fixing strategy involves marking all identified non-overflow computations 
as \texttt{unchecked}. This approach preserves the original semantics because either:
1)	the identified computation can never overflow or underflow in any execution, or
2)	if the computation does overflow or underflow, the program triggers an exception due to a check performed on a related computation.
For example, given that $a \ge b$, for an identified calculation 
like $b+k$, our algorithm ensures that a related computation, $a+k$, 
within the same function is always checked. If $b+k$ overflows, 
an overflow exception must be triggered by $a+k$.



The remaining checker identifies cases where a stack 
variable is declared inside a loop (\textit{Alloc-Loop}).
Its implementation is relatively simple, so we’ll skip the detailed descriptions here. The fix strategy 
is to move the variable declarations outside the loop, 
as illustrated in Figure~\ref{fig:loop-alloc}.

\begin{table*}[t]
\centering
\footnotesize
{
\mycaption{tab:checker-res}
{Detected Gas Wastes.}
{
\textit{($x_y$: x real gas wastes and y false positives. SL: Slither, PO: python-solidity-optimizer, GS: GasSaver, and MM: MadMax. ``-'': zero detection.)}
}
\vspace{0.05in}
\begin{tabular}{|l|c|c|c|c|c|c|c|c|c|c|c|c|c|}

\hline
\multirow{2}{*} {} & \multicolumn{7}{c|}{\textbf{PeCatch}} & \multirow{2}{*}{\textbf{SL}}  & \multirow{2}{*}{\textbf{PO}}  & \multirow{2}{*}{\textbf{GS}} & \multirow{2}{*}{\textbf{MM}} & \multirow{2}{*}{\textbf{solc}} & \multirow{2}{*}{\textbf{LLVM}}  \\ \cline{2-8}
 & {\textbf{Uncheck}} & {\textbf{Alloc-Loop}} & {\textbf{Loop-Inv}} & {\textbf{Re-Sload}} & {\textbf{Bool}} & {\textbf{Calldata}} & {\textbf{Total}} & & & & & & \\ \hline \hline

OpenZeppelin  & $28_{0}$  & $22_0$ & $23_0$ & $5_0$ & $1_0$ & $60_0$ & $139_{0}$ & $3_{1}$ & $0_6$ & $53_1$ & - & - &  $23$\\
Uniswap V3    & - & $2_0$ & $1_0$ & $5_0$  & - & - & $8_0$ & -       & $0_1$ & -     & - & 1   &  $1$ \\
uniswap-lib   & - & $4_0$ & - & - & - & - & $4_0$ & -       & $2_2$ & -     & - & -   &  -\\
solmate       & - & - & $2_0$ & - & - & - & $2_0$ & -       & -     & -     & - & -   &  $2$\\
Seaport       & - & $42_0$ & $3_0$ & $3_0$ & - & $2_0$  & $50_{0}$ & -       & -     & $2_0$ & - & 2   &  $3$\\
\hline
SeaDrop       & -  & $1_0$ & - &  $1_0$  & $1_0$ & $2_0$ & $5_0$ & -       & -       & $2_1$ & - &  -  &  - \\
V3-Periphery  & -  & $36_0$ & $9_0$ & $9_0$ & - & $5_0$ & $59_0$ & $1_4$   & $2_3$   & $6_0$ & - & -   &  $9$ \\
V2-Periphery  & -  & $24_0$ & $4_0$ & -   & - & $4_0$ & $32_0$ & $4_0$   & -       & $0_1$ & - & -   &  $4$ \\
Uniswap V2    & -  & -  & - & $3_0$  & - & -  & $3_0$ & -       & -       & -    & - & 2  &   - \\
\hline
\textbf{Total}  & $28_{0}$ & $131_{0}$ & $42_0$ & $26_0$ & $2_0$ & $73_0$ & $302_{0}$ & $8_{5}$  & $4_{12}$ & $63_{3}$ & - & 5  & $42$ \\

\hline
\end{tabular}}
\end{table*}

\subsubsection{Memory}
One checker in this category detects loop-invariant reads inside a loop, by first
identifying reads reading the same address in different loop iterations 
and then filtering out cases where a write is performed on the address in the loop (\textit{Loop-Inv}). 
We only report reads whose memory slots are on the storage, 
the calldata, 
or the memory. 
The recommended fix for these wastes is to store 
the read result on the stack outside the loop 
and read the stack variable instead inside the loop, as illustrated in Figure~\ref{fig:loop-invar}.


\subsubsection{Storage}
Two checkers aim to pinpoint gas wastes caused by misusing 
the storage area.
The first checker identifies Boolean contract fields that occupy an entire
storage slot (\textit{Bool-Field}). When such a Boolean field is changed from \texttt{false} to 
\texttt{true}, it incurs a gas cost of more than 10K.
However, we cannot blindly report every Boolean field,
since some Boolean fields may be packed together with adjacent fields into one storage slot.
Instead, given a contract, we first analyze all its fields and fields from inherited
contracts following their declaration order 
to determine whether each field shares storage slots with its adjacent fields. 
We then report Boolean fields occupying an entire storage slot. 
As discussed in Section~\ref{sec:study}, to patch these wastes, we can convert each detected field 
to \texttt{uint256} and use 1 and 2 to represent \texttt{false} and \texttt{true}, respectively. 
One such patch example is shown in Figure~\ref{fig:bool}.

The second checker pinpoints instances where reading a storage variable with \texttt{sload} could be avoided by utilizing a stack variable (\textit{Re-Sload}).
Specifically, we target two gas-inefficient code patterns: 
1) two consecutive reads are made on the same storage variable
without any writes to the variable in between; 
and 2) a write to a storage variable is followed by a read on the same variable. 
In both cases, the second read to the storage 
could be avoided by storing the data on the stack, 
as shown by Figure~\ref{fig:sstore-sload}.
We perform live variable read/write analysis to detect these two patterns.

\subsubsection{Calldata}
One checker detects instances (\eg, Figure~\ref{fig:calldata}) 
where a \texttt{memory} function argument
can be changed to \texttt{calldata} to save gas (\textit{Calldata}).
We consider both functions only 
accessed from outside the contract,
and functions called by functions within the same contract. 
For each formal parameter labeled as \texttt{memory}, we inspect
whether the function modifies its memory area. 
If the function is called by any function within the same contract, we further
inspect whether the corresponding real parameter is on the calldata. 
We perform an iterative inspection on all functions, until we no longer find any parameters
that can be changed to \texttt{calldata}.
\subsection{PeCatch Evaluation}
\label{sec:exp}
%
%
\subsubsection{Methodology}
We implement PeCatch using static analysis 
framework Slither~\cite{slither}.
Slither takes Solidity source code as its input and 
converts it into SSA form, aiding researchers in building
static detectors for bugs and vulnerabilities. 
In sum, PeCatch contains 2065 lines of Python code, 
encompassing static analysis routines utilized by various checkers 
(\eg, identifying loops), 
as well as waste detection algorithms tailored for each checker.

\italicparagraph{Benchmarks.}
Besides the programs discussed in Section~\ref{sec:study}, 
we further include four additional Solidity programs in our evaluation to
ensure our checkers 
can gas wastes beyond the examined applications. 
We focus on programs that are mature, with a long commit history, 
popular (evidenced by a significant number of GitHub stars), 
and widely used within the community. 
For example, Uniswap V2 has over 3.1K stars on GitHub and 
has been forked more than 3.1K times. Moreover, 
the source code size of the selected programs ranges from 424 lines to 4,053 lines, 
which is comparable to the programs previously studied. We believe these nine benchmarks are sufficient to evaluate PeCatch, 
as they collectively 
contain 36,892 lines of source code and 584 contracts, providing a statistically confident basis 
for evaluation, and reflect common use cases of 
the Solidity programming language. Table~\ref{tab:checker-res} lists all the benchmarks used in our experiments.


\italicparagraph{Research Questions.}
We aim to answer the following research questions:

\begin{itemize}[noitemsep, topsep=0pt, leftmargin=.25in]
\item {\textit{Effectiveness:}} How well does PeCatch perform in detecting previously 
unknown gas wastes in real-world Solidity programs? 


\item {\textit{Benefits:}} How much gas can be saved by fixing 
the detected wastes?

\item {\textit{Coverage:}} What percentage of real-world gas wastes can PeCatch identify?

\end{itemize}

Specifically, we run PeCatch on the latest versions of the selected programs, 
counting the number of detected gas wastes and false positives to assess 
its effectiveness. We manually patch the detected wastes 
and execute all unit tests in the benchmark programs to measure 
the amount of gas and gas fees saved. Since the exact number of gas wastes 
in the latest benchmark versions is unknown, we cannot use them 
to evaluate PeCatch's coverage. Therefore, we rely on the wastes 
analyzed in Section~\ref{sec:study}. We manually inspect these wastes and count 
how many of them PeCatch can identify to assess its coverage.

\italicparagraph{Baseline Techniques.}
We compare PeCatch with six baseline techniques, 
including four static gas-waste detection techniques: Slither~\cite{slither}, 
python-solidity-optimizer~\cite{brandstatter2020characterizing}, 
GasSaver~\cite{gassaver}, and MadMax~\cite{grech2018madmax} 
(referred to as SL, PO, GS, and MM, respectively);
and two compiler optimization suites: 
the Solidity compiler (solc)~\cite{solidity-optimizations} and 
LLVM compiler optimizations~\cite{llvm-analysis}.
Due to compatibility issues, the original code of MadMax cannot work on 
recent Solidity programs. Thus, we reimplement its detection algorithms using Slither.
Furthermore, since LLVM does not support Solidity, 
we carefully review LLVM's documentation 
to understand the compiler optimization algorithms it employs. For promising algorithms, we conduct a deep inspection of their implementation 
to assess whether each gas waste detected by PeCatch could be pinpointed. Additionally, we run solc on each gas waste identified by PeCatch and 
examine the assembly code after optimization to 
determine whether solc can identify the waste.

\italicparagraph{Experimental Setting.}
All our experiments are conducted on
a Mac Pro Notebook, with an Intel Core i7 CPU, 16 GB RAM, and MacOS Ventura 13.4.1 (c).

\subsubsection{Effectiveness}
\label{sec:effectiveness}

As shown in Table~\ref{tab:checker-res},
PeCatch reports 302 suspicious code sites. We carefully examine these sites 
and confirm all of them are previously unknown gas wastes, with 203 
from the five studied programs 
and 99 from the four additional programs. 
Our assessment involves checking whether 
each code site follows the corresponding code pattern of the checker, 
assessing whether we can design a patch that effectively reduces gas consumption, 
and validating the results of all available unit tests.

Each of our checkers detects 
some previously unknown gas wastes. 
Among them, Alloc-Loop pinpoints the most gas wastes, 
uncovering 131 wastes from seven benchmark 
programs. The potential reason is that programmers may be 
unaware that the Solidity compiler does not optimize allocations inside loops as traditional programming language compilers do.
Bool-Field detects fewest gas wastes, 
pinpointing only two instances from OpenZeppelin
and SeaDrop.

\italicparagraph{Gas Waste Reporting.}
We try our best to report the detected gas wastes to the programmers. 
For 63 of the detected wastes, the buggy code sites were removed due to 
code refactoring that occurred before we completed our experiments. 
As a result, we did not report these wastes.
We created pull requests for all other 239 gas wastes 
and submitted the pull requests to the corresponding GitHub repositories. 
So far, programmers have fixed eight of them and 
mentioned that they plan to address another 32 in future versions.
For 56 of the reported wastes, the programmers have confirmed the buggy code snippets indeed consume more gas than our proposed patches. However, they decide not to merge our pull requests. 
This includes six cases where our pull requests 
are approved by at least one programmer but have not been merged yet, 23 cases where the programmers consider the gas savings to be marginal, 
26 cases where they are concerned that code readability might be negatively affected 
after adding \texttt{unchecked}, and one case where the 
contract 
containing the waste is scheduled to be deprecated.
We have not received any responses from programmers 
regarding the remaining 143 wastes.

\italicparagraph{Baseline Comparison.}
PeCatch detects 4.8\,$\times$ to 75.5\,$\times$ more gas wastes than SL, PO, 
and GS, as shown in Table~\ref{tab:checker-res}. 
Unlike these techniques, PeCatch targets gas-waste patterns 
related to Solidity’s unique language features. Wastes following these patterns 
are prevalent in real-world Solidity programs, 
resulting in PeCatch's higher detection capability.
Even for the same type, PeCatch identifies a greater quantity. 
For instance, PeCatch, SL, and GS all attempt to pinpoint 
parameters that can be switched from \texttt{memory} to \texttt{calldata}. 
However, PeCatch identifies 73 wastes, 
compared to just four detected by SL and 58 detected by GS. 
PeCatch detects more because it also 
considers functions called by another function in the same contract.
MM fails to pinpoint any gas wastes, 
because MM focuses on contract fields that are dynamic arrays, 
but dynamic arrays are rarely used  
as contract fields in recent Solidity programs.
Additionally, as shown by the subscripts in Table~\ref{tab:checker-res},
PeCatch reports zero false positives across all programs.
In contrast, SL, PO, and GS all have false positives, with
false-positive rates (number of false positives over true positives) of
62.5\%, 300\%, and 4.8\%, respectively. 
Those false positives are due to inaccurate analysis algorithms,
or the difficulty in fixing the detected wastes. 

Existing compiler optimization algorithms also fail to detect most gas wastes identified by PeCatch. 
Solc's algorithms can optimize only five wastes resulting from consecutively 
reading a contract field twice, without any intervening instructions. 
PeCatch's Re-Sload not only identifies these five wastes 
but also pinpoints an additional 21 wastes involving more complex code. 
LLVM's algorithms identify 42 wastes that are 
detected by PeCatch's Loop-Inv. 
\textit{In sum, PeCatch is much more effective than
existing gas-waste detection techniques and gas-optimization techniques.}

\italicparagraph{Execution Time.}
We measure the execution time of PeCatch on each program, by running
each checker \textit{10 times} and summing up the average 
execution time of each checker. 
Overall, it takes PeCatch from 4 seconds to 7 minutes to analyze a benchmark program. Thus, \textit{PeCatch’s can potentially be used during Solidity 
programmers’ daily development practice.}

\begin{tcolorbox}[size=title]
{\textbf{Answer to Effectiveness:} 
PeCatch pinpoints a large number of gas wastes in real Solidity programs 
within a short time frame and reports zero false positives; it outperforms existing 
techniques in both the quantity of detected wastes and accuracy.
}
\end{tcolorbox}

\subsubsection{Benefits}
\label{sec:benefits}

We employ two methods to estimate the amount of gas saved after fixing the detected wastes, along with their monetary impact. Section~\ref{sec:benefits}

The first method utilizes all the unit tests from the programs listed in Table~\ref{tab:checker-res}. 
We manually patch all detected wastes using the fixing mechanisms 
outlined in Section~\ref{sec:pecatch}. Fixing these wastes 
requires a basic understanding of Solidity, and each patch takes 
approximately one to three minutes to implement, as the fixing mechanisms are relatively simple. These are demonstrated in 
Figure~\ref{fig:unchecked}, and Figures~\ref{fig:loop-alloc}--\ref{fig:calldata}.
After applying the fixes, we run all unit tests and compare 
the gas usage between the original versions of the programs 
and the versions with all wastes resolved. The gas usage is 
reduced by 223,382 units in \textit{one} single execution of all 
tests 
for the patched versions. This reduction corresponds to a cost savings of \$46.51.

The second method leverages the transaction traces collected 
in Section~\ref{sec:chain} to analyze how frequently the gas-waste patterns identified by PeCatch occur in practice. 
In the beginning, we figure out the opcode sequences associated with 
these gas-waste patterns. Then, we search for these sequences within 
the transaction traces. In the end, we calculate the potential gas savings if the wastes 
were fixed. Overall, the gas saved represents 4.55\% of the total gas consumption, 
amounting to approximately \$0.73 million per day.

\begin{tcolorbox}[size=title]
{\textbf{Answer to Benefits:} 
Gas wastes covered by PeCatch consume a significant amount of gas 
and result in substantial financial costs in practice, 
making it highly beneficial to resolve them.

}
\end{tcolorbox}

\begin{table}[t]
\centering
\footnotesize
\mycaption{tab:coverage}
{Gas Waste Coverage.}
{
\textit{(``+8'': eight more gas wastes can be resolved after fixing the compiler issues identified by us. SL: Slither, PO: python-solidity-optimizer, GS: GasSaver, and MM: MadMax. )}
}
\vspace{0.1in}
{
 \setlength{\tabcolsep}{1.4mm}{
\begin{tabular}{|l|c|c|c|c|c|c|c|c|}
\hline
\multirow{2}{*}{\textbf{\makecell{Root Cause\\ (Store Area)}}} &\multirow{2}{*}{{\textbf{Wastes}}} & \multicolumn{7}{c|}{\textbf{Techniques}}        
\\ \cline{3-9}
 & & {{\textbf{PeCatch}}} & {{\textbf{SL}}} & {{\textbf{PO}}}  &  {{\textbf{GS}}} &  {{\textbf{MM}}} &  {{\textbf{solc}}} &  {{\textbf{LLVM}}} \\ \hline \hline

\textbf{Stack}          & $24$ & $7+8$      & $0$ & $0$  & $0$  & $0$ & $0$ & $0$       \\ \hline
\textbf{Memory}         & $4$  & $3$       & $3$ & $0$  & $0$  & $0$ & $0$ & $3$        \\ \hline
\textbf{Storage}        & $20$ & $18$      & $0$ & $0$  & $0$  & $0$ & $3$ & $0$        \\ \hline 
\textbf{Calldata}       & $6$ & $5$        & $5$ & $0$  & $5$  & $0$ & $0$ & $0$        \\ \hline 
\hline
\textbf{Total}          & $54$ & $33+8$      & $8$ & $0$  &$5$   & $0$ & $3$ & $3$         \\ \hline

\end{tabular}
}
}

\end{table}

\subsubsection{Coverage}
We use the gas wastes studied in the empirical study (Table~\ref{tab:root}) 
as the ground truth dataset. We manually examine each waste and 
count how many can be detected by each technique to evaluate their coverage 
of gas wastes. As shown in Table~\ref{tab:coverage}, PeCatch detects 33 out 
of the 54 gas wastes, demonstrating \textit{its robust coverage of real gas wastes 
associated with Solidity’s unique features.} 
In contrast, the baseline techniques detect significantly fewer gas wastes. 
Among the baselines, SL identifies the most, but it can only detect eight gas wastes. 
GS detects five wastes, which are associated with \texttt{memory} function parameters 
that can be changed to \texttt{calldata}. Each of the two 
compilers (solc and LLVM) detects three wastes. 
The three wastes identified by solc are caused by consecutive accesses to 
the same contract field, while three wastes detected by LLVM are 
caused by accessing the same memory location within a loop. 
PO, based on traditional compiler optimization algorithms, 
and MM, targeting gas wastes related to contract fields that are dynamic arrays, fail to detect any gas wastes in our ground-truth dataset.

PeCatch misses 21 wastes for the following reasons. 
First, the Uncheck checker overlooks five wastes because programmers rely on workload information to uncheck the computations, 
which PeCatch is unaware of. 
Second, the same checker misses two additional wastes 
due to its limited effectiveness in path-condition analysis, 
causing it to miss non-overflowable cases ensured by complex path conditions. 
Third, Re-Sload 
misses two gas wastes as it does not conduct inter-procedural analysis. Fourth, Calldata-Para
misses one waste as it fails to consider scenarios where a return can be changed from \texttt{memory} to \texttt{calldata}. 
Fifth, we consider using \texttt{\&\&} in an \texttt{if}’s condition and 
returning a local variable are due to bugs in the Solidity compiler 
and do not implement 
checkers for these cases. 
 Lastly, PeCatch misses the remaining three wastes 
because it does not consider specific opcodes 
(\eg, \texttt{mstore8}) or particular values (\eg, \texttt{type(uint256).max}) in Solidity.
To avoid these misses, we plan to extend PeCatch by conducting more complex static analysis (\eg, interprocedural analysis,
path-condition analysis) and considering more Solidity language features in its future versions.

\begin{tcolorbox}[size=title]
{\textbf{Answer to Coverage:} 
PeCatch is capable of detecting a significant portion of gas wastes 
associated with Solidity’s unique language features.
}
\end{tcolorbox}


\subsection{Identifying Issues in the Solidity Compiler}

We design toy programs to reproduce the wastes due to compiler
implementation issues discussed in 
Section~\ref{sec:study} and Section~\ref{sec:chain}.
Then, we identify which part of the compiler causes the issues 
and examine why the compiler programmers make the mistakes.
Through this process, we identify two compiler errors that cause gas wastes when using \texttt{\&\&} in an \texttt{if}'s condition and when returning \texttt{retLocal} in Section~\ref{sec:study},
and 13 errors for the 13 gas-inefficient opcode sequences 
in Section~\ref{sec:chain}.
However, one sequence in Section~\ref{sec:chain} involving 
the use of \texttt{||} in an \texttt{if}'s condition shares the same compiler code as
using \texttt{\&\&} in an \texttt{if}'s condition. Thus, we detect a total of 14 distinct 
compiler issues. 
We report all of them to the Solidity team.
However, the programmers have decided not to address the 
issues, as all of them are in the legacy pipeline, which is 
set to be replaced by the IR pipeline.

Most of the issues (11 out of 14) arise from using a generic function for various 
scenarios, missing optimization opportunities for specific cases. An example of 
this is the issue associated with using \texttt{\&\&} in an \texttt{if}'s condition in Section~\ref{sec:not-optimized}. 

\italicparagraph{Gas Saving.}
Using the collected transaction traces, we estimate that fixing these compiler errors 
would save approximately \$0.03 million in gas fees every day. 
This highlights \textit{the monetary impact of the compiler issues.}

\section{Limitations and Discussion}

\italicparagraph{Threats to Validity.}
Potential threats to validity include the representativeness of the selected applications for both the study and evaluation, the representativeness of the gas wastes studied, 
and the methodology used in our gas-waste study.

Regarding application representativeness, our study selects five popular 
and widely used applications that are representative of common 
use cases for the Solidity programming language. However, 
there may be additional use cases not covered.
For the evaluation, we select four additional applications based on the same criteria. 
Although the total evaluated applications contain 36,892 lines of source code 
and 584 contracts, there may still be uncovered code patterns, 
which could lead to potential false positives and false negatives.

All of the gas wastes we studied are sourced from GitHub commit history. 
We believe our collection method captures all significant gas wastes reported 
through GitHub issues, as any time a programmer addresses 
a reported issue, a corresponding commit is made. 
However, since we collect gas wastes through keyword searches, 
we acknowledge that some gas wastes may have been fixed but not 
documented using the relevant keywords, and therefore are missed in our search. 
Additionally, some gas wastes may not have been recognized by users or programmers, 
and as a result, are not patched. While these unresolved bugs may have different 
root causes than the fixed ones, we believe they are likely less significant than the fixed bugs included in our study.
It is also possible that programmers use findings from previous papers to prevent gas wastes in certain old patterns from occurring again.

In terms of our examination methodology, we analyze the source code, 
textual descriptions, and discussions among programmers for each gas waste. 
Each gas waste is reviewed by at least two authors of the paper, 
both of whom have substantial knowledge of Solidity. 
Any disagreements are resolved through multiple rounds of discussion. 
However, we acknowledge that the results are influenced by our 
personal expertise. Additionally, we did not measure the initial degree of agreement between the two authors, which means potential issues in the examination process may not have been exposed.

\italicparagraph{Discussion.}
While manually fixing the wastes detected by PeCatch does not require specialized knowledge, developing an automated technique to generate patches for the identified wastes would be highly beneficial for PeCatch users. We plan to build this technique in future work.
Large language models (LLMs) like ChatGPT have recently been explored for tasks such as code generation and bug fixing. 
It is promising to apply LLMs to identify gas wastes and optimize gas usage. 
However, LLMs are known to have limitations in prompt length, so the input 
Solidity code must be appropriately segmented. 
Additionally, LLMs may suffer from hallucinations and are highly 
dependent on effective prompting techniques. The gas wastes we studied can serve as one-shot examples to help LLMs better understand the issues and optimizations, thus reducing the likelihood of hallucinations.
\section{Conclusion}

Facing the growing popularity of blockchain systems and smart contracts, 
this paper presents an empirical study on real-world gas wastes 
in Solidity smart contracts and on-chain transaction traces. 
The study inspects three key aspects: where gas wastes occur, 
why they are not optimized away by the compiler, 
and their potential solutions, with particular focus on gas wastes 
related to Solidity's unique language features. 
To demonstrate the utility of our findings, we build a static 
gas-waste detection technique for Solidity source code 
and identify implementation errors in the Solidity compiler causing gas wastes. 
We anticipate this research will enhance our understanding of gas wastes and encourage further research and practical efforts to mitigate them.

\section*{Acknowledgment}

We sincerely thank the reviewers for their detailed and invaluable feedback.
Shihao Xia, Mengting He, and Linhai Song were supported 
by NSF grants CNS-1955965 and CCF-2145394.
Tingting Yu was supported by NSF grant CCF-2402103.
Nobuko Yoshida was supported by EPSRC grants EP/T006544/2, EP/T014709/2, EP/Y005244/1, EP/V000462/1, and EP/X015955/1, and 
EU Horizon (TARDIS) 101093006.

\clearpage
\bibliographystyle{IEEEtran}
\bibliography{gas}


\begin{IEEEbiography}[{\includegraphics[width=1in,height=1.25in,clip,keepaspectratio]{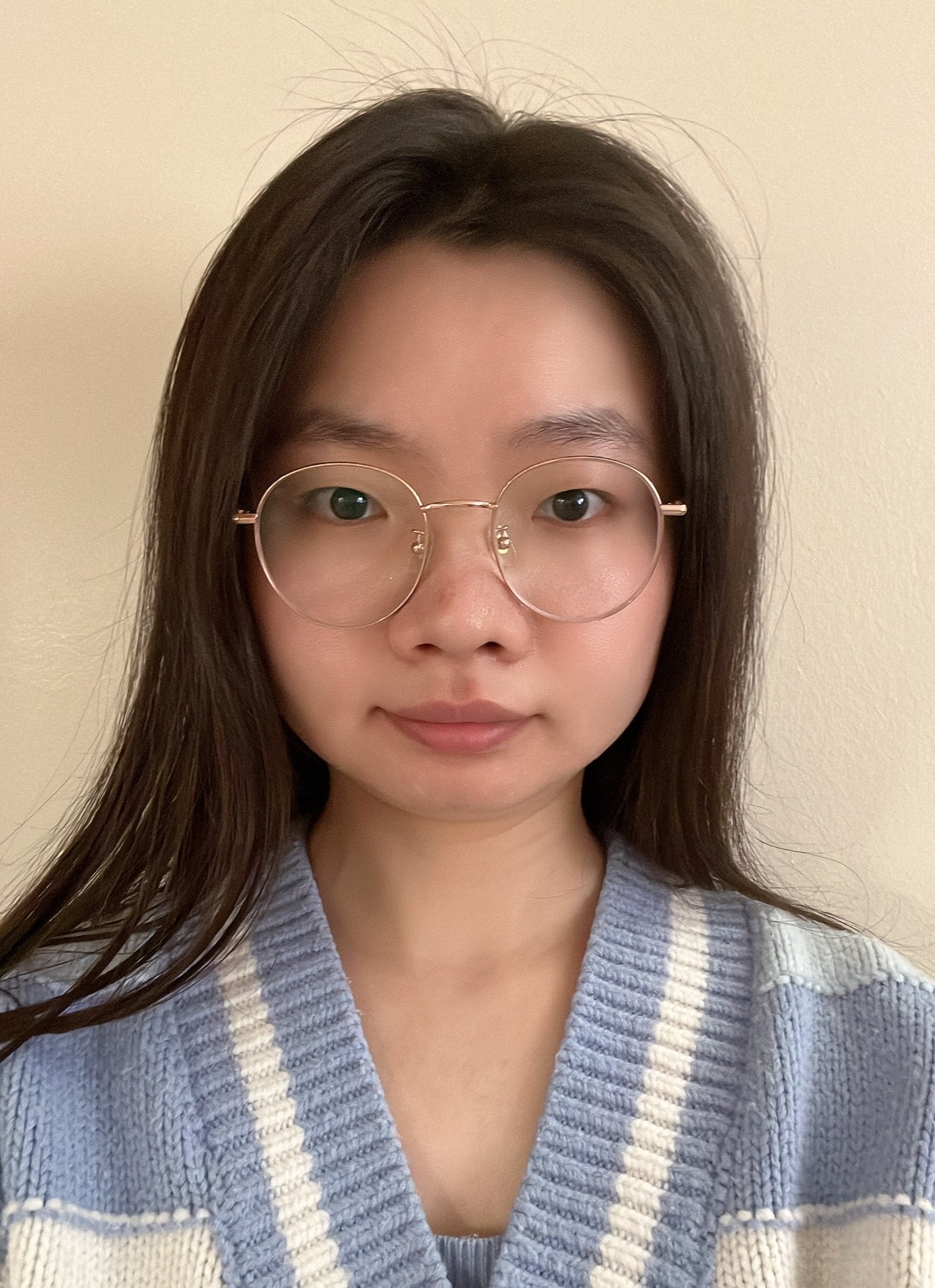}}]{Mengting He} is a Ph.D. student in the College of Information Sciences and Technology at the Pennsylvania State University. Her main research interests focus on program analysis and software reliability. She received her bachelor's and master's degrees from Nanjing University.

\end{IEEEbiography}

\begin{IEEEbiography}[{\includegraphics[width=1in,height=1.25in,clip,keepaspectratio]{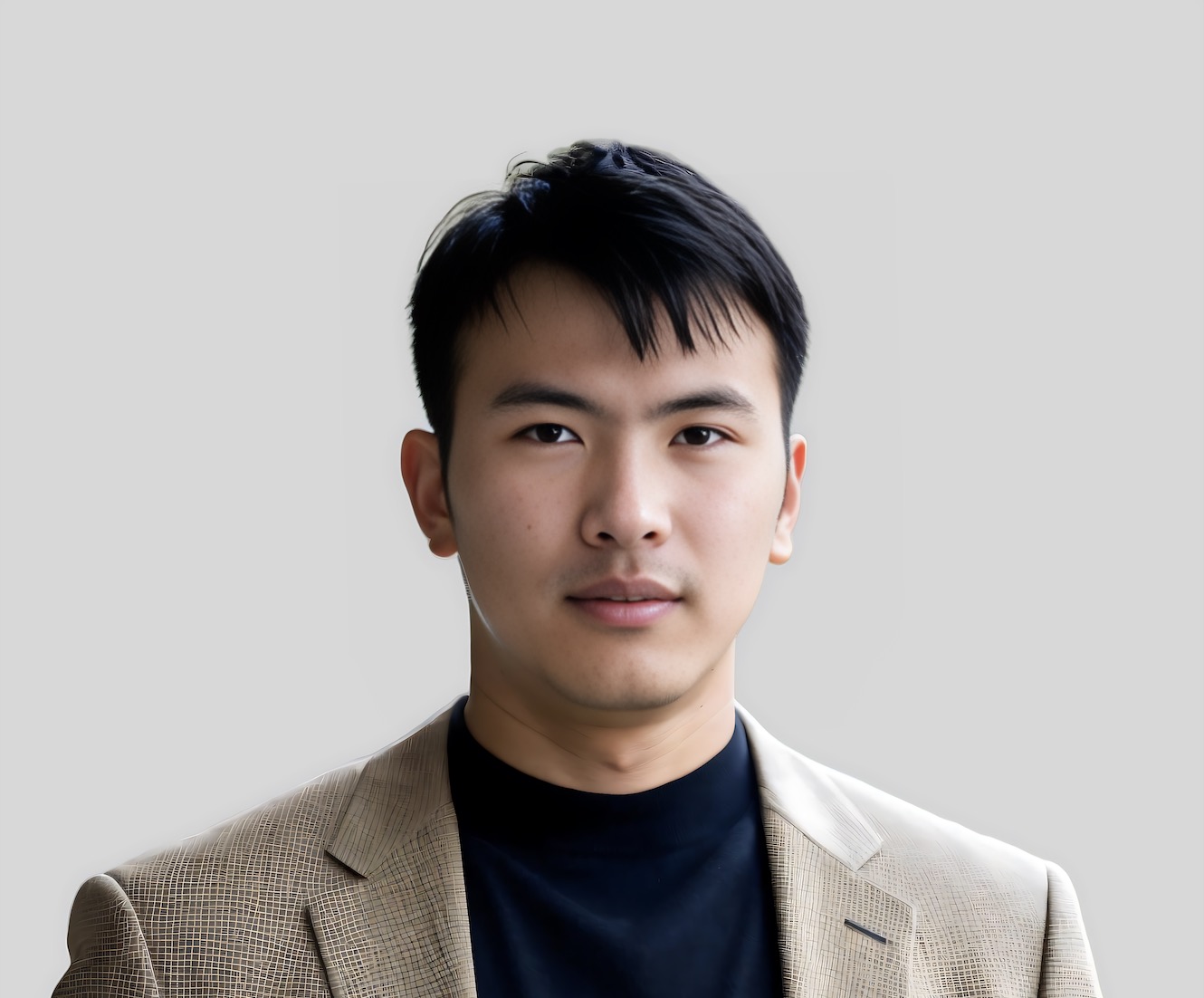}}]{Shihao Xia} is a Ph.D. student in the College of Information Sciences and Technology at the Pennsylvania State University. His main research interests focus on program analysis and software reliability. He received his bachelor's degree from Worcester Polytechnic Institute. 

\end{IEEEbiography}

\begin{IEEEbiography}[{\includegraphics[width=1in,height=1.25in,clip,keepaspectratio]{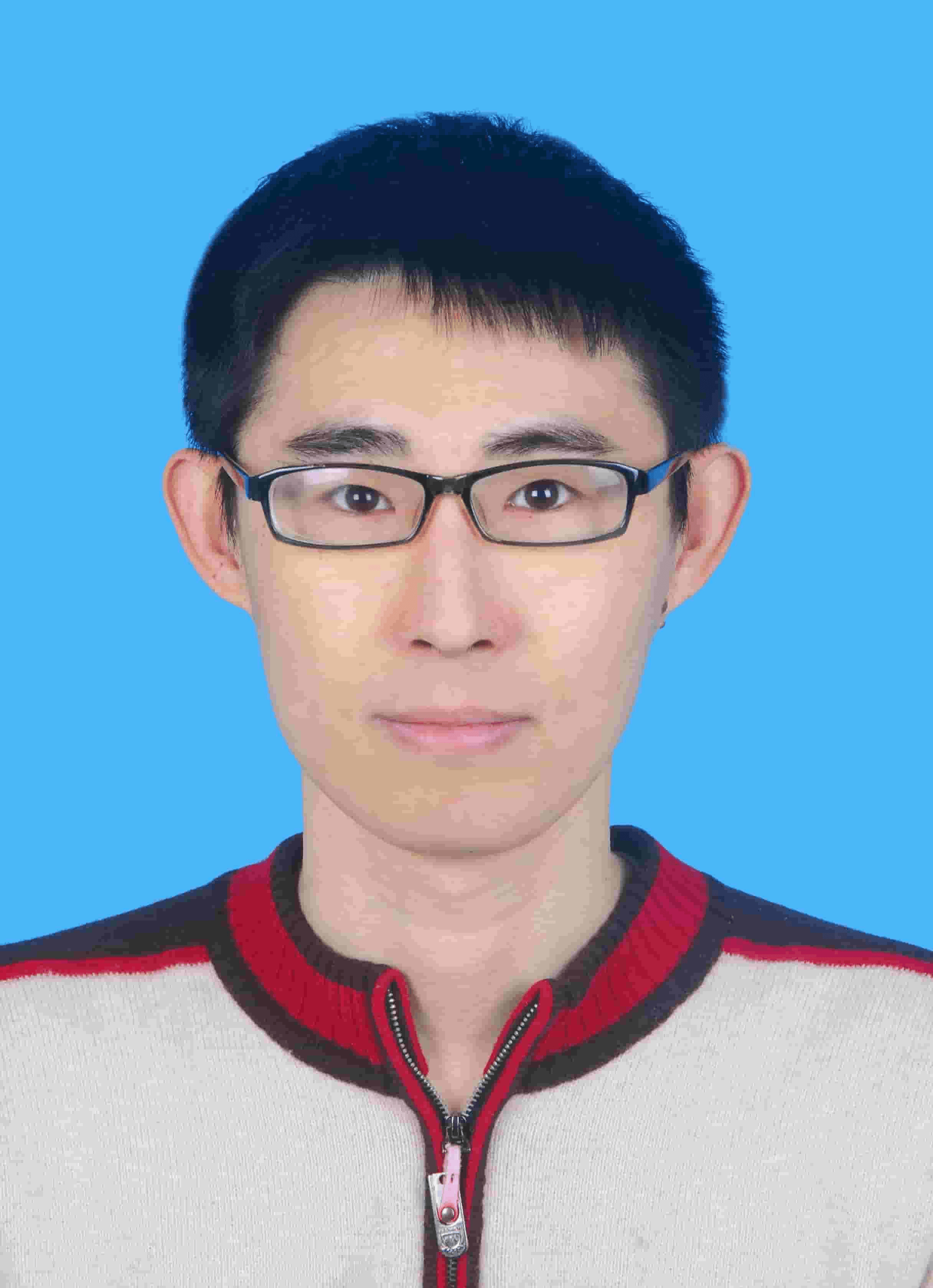}}]{Boqin Qin}
received his Ph.D. in computer science from Beijing University of Posts and Telecommunications.
	He is a researcher and developer at China Telecom Cloud Technology Co., Ltd. 
	He was a visiting student at the Pennsylvania State University.
	His current research interests focus on program analysis, blockchain security, and distributed systems.
\end{IEEEbiography}

\begin{IEEEbiography}[{\includegraphics[width=1in,height=1.25in,clip,keepaspectratio]{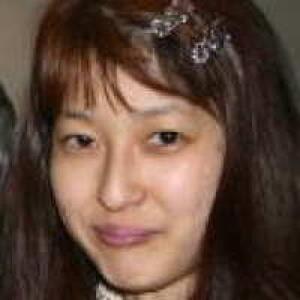}}]{Nobuko Yoshida} 
completed her Ph.D. jointly at the University of Keio and the University of Manchester. She is Christopher Strachey Chair of Computer Science in the University of Oxford. She is an EPSRC Established Career Fellow and an Honorary Fellow at Glasgow University. Her research interests focus on the theory and practice of concurrent and distributed languages and systems.

\end{IEEEbiography}

\begin{IEEEbiography}[{\includegraphics[width=1in,height=1.25in,clip,keepaspectratio]{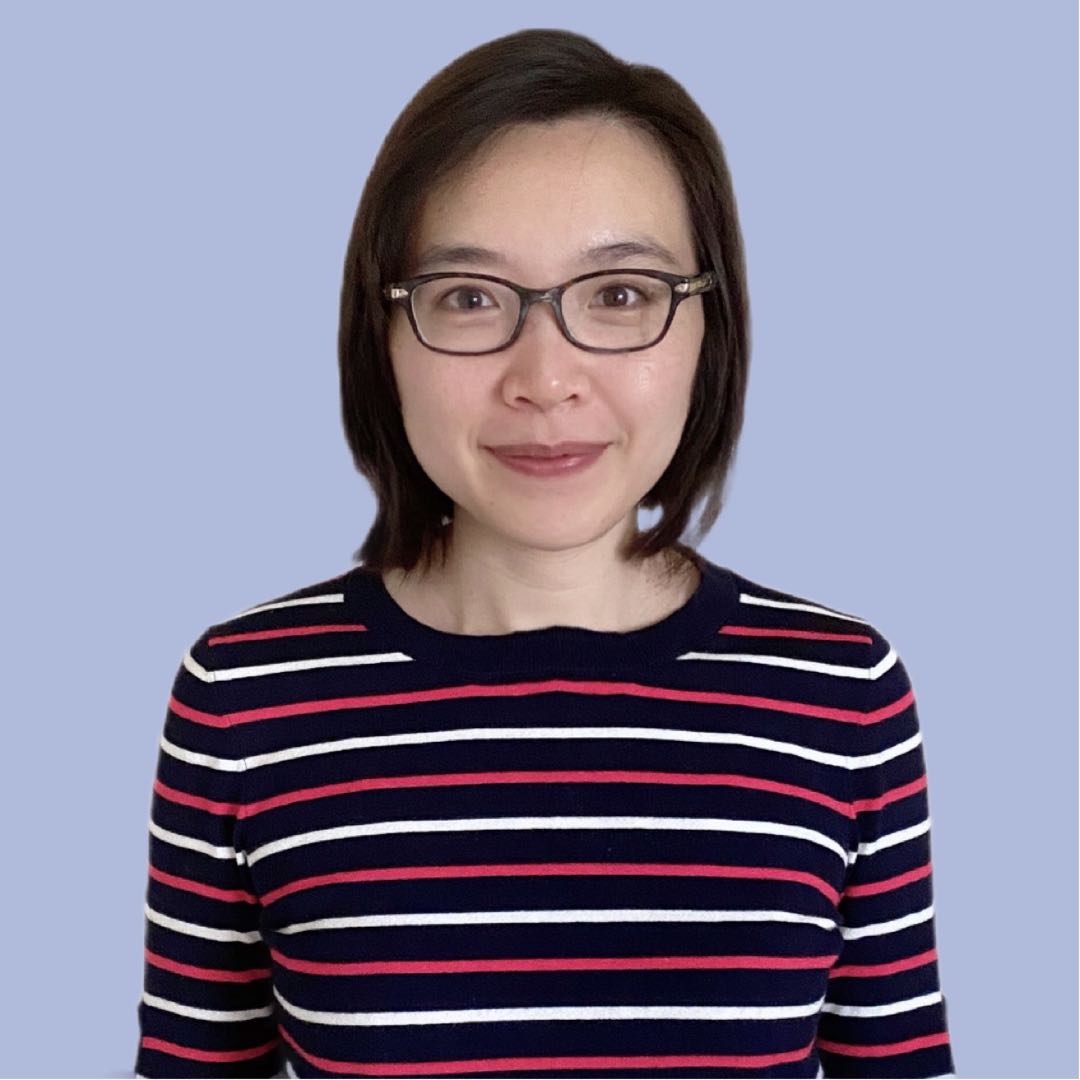}}]{Tingting Yu}
received her Ph.D. from the University of Nebraska-Lincoln. She is an associate professor in the Computer Science and Engineering Department at the University of Connecticut. Her research interests are software engineering, software testing, software architecture, AI for software engineering, and domain-specific software engineering (\eg, bioinformatics, cyber-physical systems).
\end{IEEEbiography}

\begin{IEEEbiography}[{\includegraphics[width=1in,height=1.25in,clip,keepaspectratio]{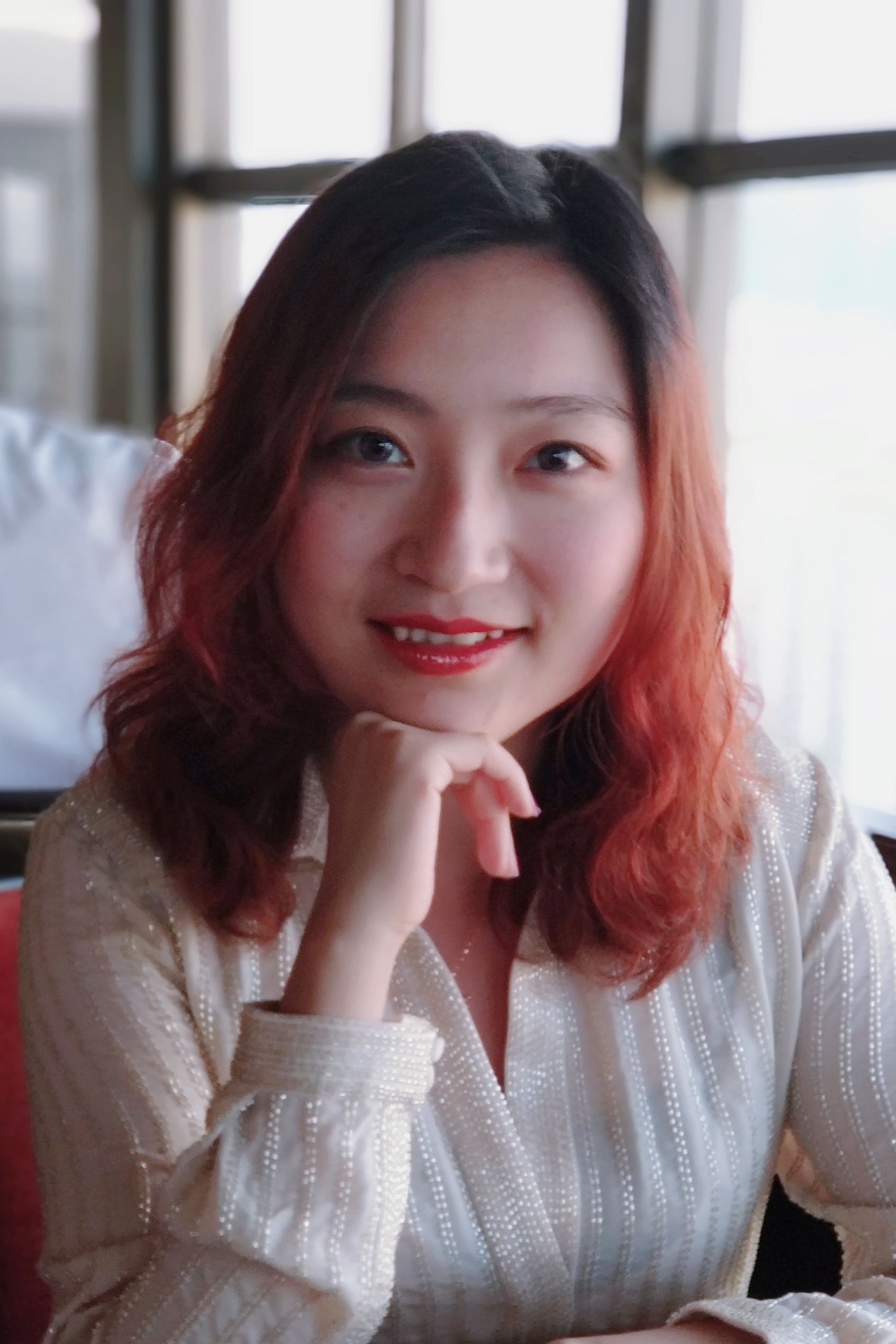}}]{Yiying Zhang}
received her Ph.D. from the University of Wisconsin-Madison. 
She is an associate professor in the Computer Science and
	Engineering Department at the University of California, San Diego. Her research interests are operating systems, distributed systems,
	computer architecture, and datacenter networking.
\end{IEEEbiography}

\begin{IEEEbiography}[{\includegraphics[width=1in,height=1.25in,clip,keepaspectratio]{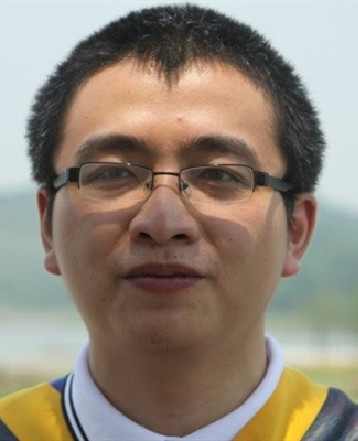}}]{Linhai Song}
received his Ph.D. from the University of Wisconsin-Madison. He is an associate professor in the College of Information Sciences and Technology at the Pennsylvania State University. His main research interests are software reliability and software systems.
\end{IEEEbiography}

\end{document}